\documentclass[onecolumn]{article}
\usepackage{psfig}
\usepackage{astron}

 \def\figOyn#1{#1} 
  \def\figPSyn#1{}   

  \def\figpsyn#1{}   

\def\figyn#1{\figOyn{#1}\figPSyn{#1}\figpsyn{#1}}

\begin{document}

\def\gh{{\rm gh}}
\def\rg{{\rm rg}}
\def\jet{{\rm jet}}
\def\power{{\rm power}}

\def\eps{\varepsilon}
\def\CR{{\rm CR}}
\def\gal{{\rm gal}}
\def\gas{{\rm gas}}
\def\eps{\varepsilon}

\title{Isotropization of Ultra-High Energy Cosmic Ray Arrival
Directions by Radio Ghosts}
\author{Gustavo Medina-Tanco \\
Instituto Astron\^{o}mico e Geof\'{\i}sico, Universidade de S\~ao Paulo, Brazil \\
gustavo@iagusp.usp.br\\
and\\
Torsten A.  En{\ss}lin\\MPI f\"ur Astrophysik \& MPI f\"ur
Radioastronomie, Germany\\
Department of Physics, University of Toronto, Canada\\
ensslin@mpa-garching.mpg.de}
\maketitle

\begin{abstract}
The isotropy in the ultra high energy cosmic ray (UHECR) flux
observed by Yakutsk and AGASA experiments, is a very strong
constraint to production and propagation models alike.  Most of
the scenarios proposed in the literature should produce a sizable
anisotropy as either extragalactic luminous or dark matter is
normally associated with the invoked particle sources.  We explore
the possibility that the magnetic fields in fossil cocoons of
former radio galaxies -- so called {\it radio ghosts}
\cite{Ringberg99}-- are able to scatter UHECR in the intergalactic
medium giving rise to the observed isotropy. We show, through
numerical simulations, under which conditions this process can be
operative and the magnitude of the effect. We further demonstrate,
that if radio ghosts mix with the ambient medium, they might be
able to produce the observed magnetic fields in clusters of
galaxies. In the case of mixing, the UHECR isotropization would be
even stronger than in our conservative estimates.
\end{abstract}

\tableofcontents


\section{Introduction}
\subsection{The Problem of UHECR Isotropy}

The upper end of the cosmic ray spectrum, at total energies above
the (as yet unobserved) Greisen-Zatsepin-Kutzmin (GZK) cut-off
($\sim 4 \times 10^{19}$ eV - \cite{GZK_G66}, \cite{GZK_ZK66})
represents a challenge
for particle-physicists and astrophysicists alike. The nature
of the sources of these ultra-high energy cosmic rays and their
distance scale are still unknown. Only our own galactic disk can
be ruled out at present as a major source site, as a compatible
anisotropy has not been observed by any of the experiments sensitive
to the UHECR energy range
\cite{AG_1e19,Bird98,GMT_AAW_DM_halo}.

Although photons, neutrinos, or some unknown particle cannot be
disregarded, the muon to electron ratio measured for extensive air
showers \cite{Halzen95,AGASA_pairs} points to hadrons as the
primaries hitting the upper atmosphere and triggering the
cascades. Neutrons with relativistic factor $\gamma \sim 10^{11}$
decay into protons after a path of $\sim 1$ Mpc. Heavy nuclei, on
the other hand, may lose as much as $\sim 2-4$ nucleons per Mpc
\cite{Puget76,Cronin92} due to photodisintegration in interactions
with the cosmic microwave and infrared backgrounds. Therefore,
although new estimates of the infrared background
\cite{Malkan_and_Stecker_98} seem to indicate that the previous
photodisintegration rates may be overestimated by a factor of $10$
\cite{Stecker98}, UHECR, unless galactic, are very likely light
nuclei, probably mainly protons. We will assume the latter in the
remaining of this work.

Depending on the large scale configuration of the intergalactic
magnetic field (IGMF), field values as low as $B_{IGMF} \sim
10^{-9}$ G \cite{Kromberg94}  may be expected. Consequently,
proton gyroradius at $E\sim 10^{20}$ eV can be of the order of
$10^{2}$ Mpc. This represents a potentially interesting
perspective for astronomy since UHECR should, under favorable
conditions, point to their sources opening a new astronomical
window. This would allow not only the direct study of their
sources, but also the measurement of the intervening intergalactic
and galactic (particularly halo) magnetic fields.

Relevant to this potential is the nature of the production
mechanism involved. So far, both bottom-up and top-down UHECR
production mechanisms have been devised and constitute perfectly
viable scenarios. The first set of models invoke some kind of
acceleration agent that is always spatially associated with
baryonic matter, closely traced by luminous matter. Top-down
mechanisms, on the other hand, resort to the decay of relics from
GUT phase transitions in the early Universe \cite{Sigl_TD}. Under
the assumption that relics aggregate like dark matter (DM) and
that their decay products are susceptible of inelastic
interactions with the CMB, then any observed anisotropy should
trace the nearby distribution of DM. Therefore, the UHECR flux
should have two components, one originated inside the galactic
halo and another extragalactic still roughly traced by luminous
matter. \cite{Dubovsky_DM} showed that, under general conditions,
the halo component would dominate the extragalactic flux by at
least two orders of magnitude. This is only true, however, in the
unrealistic case of dark matter uniformly distributed in
intergalactic space. Nevertheless, dark matter aggregates strongly
and tends to be overabundant, by factors of $\sim 10^{2}$, in the
center of galaxy clusters when compared to its abundance in the
halos of isolated galaxies. It can therefore be shown that
\cite{UHECRiso00}, in a sample of 47 events (as AGASA's above $4
\times 10^{19}$ eV), and assuming Virgo as the only source of
extragalactic events, 3-7 events should originate in Virgo and
arrive inside a solid angle of approximately the size of the
cluster. This could give rise to a slight anisotropy that
correlates with the SGP when combined with the almost isotropic
flux originated in a large galactic halo.

Photo-pion production interactions with the cosmic microwave
background (CMB) restrict the source region to nominally $D
\stackrel{~}{<} 100$ Mpc \cite{Waxman97,Hillas98}. Therefore, the
signatures of the characteristic inhomogeneity of the local
universe should be clearly observable in the angular arrival
distribution of UHECR regardless of the production mechanism. In
spite of the low statistics available at present, the latter does
not seem to be the case \cite{AG_1e19,GMT_isotropy00}.

If UHECR are charged particles originated inside the large scale
structures in the nearby Universe, then any isotropization
must be due to intervening magnetic fields. This could happen
either inside the galactic halo, like in the presence of a
magnetized galactic wind \cite{Ahn_etal_GW00}, by the interaction
with highly structured IGMF inside walls and filaments
\cite{GMT_APJL98_IGMF} or by the scattering off magnetic
irregularities permeating the intergalactic medium.

In the present work we analyze the latter possibility considering
radio ghosts \cite{Ringberg99} as scattering centers of UHECR protons.

\subsection{The Existence of Radio Ghosts}
Active galaxies eject large amounts of radio emitting plasma -- short:
{\it radio plasma} -- into their environment. There it forms the
typical cocoons of radio galaxies. The radio emission results from
synchrotron emission of a population of relativistic electrons in the
radio plasma's magnetic fields. It is possible to estimate the minimal
energy density of the electron and magnetic components of the radio
plasma required in oder to produce the observed emissivity.  This
minimum is given by rough energy equipartition between electron
population and the magnetic fields. The resulting minimal pressure is
typically of the order of the environmental thermal pressure,
indicating that relatively strong magnetic fields are present.

After a cosmological short time of $10^7 - 10^8$ years the radio
luminosity of the cocoon decays strongly due to radiation and
expansion energy losses of the electron population, or since the
central engine of the radio galaxy stopped its activity.  Although
undetectable by our instruments, the radio plasma is still present
in the IGM inside a fossil radio cocoon, a so called radio ghost.
The subsequent evolution of the radio plasma is unclear, since
observationally poorly constrained. It can be expected that the
strong magnetic fields of the ghosts allow it to resist erosion by
subsonic turbulence. The fossil radio cocoon should be
kinematically decoupled from the ballistic motion of the parent
galaxy and follow mostly the flow pattern of the embedding
material. From this one would expect the ghosts to have a
cosmological distribution comparable to that of the galaxies. And
one would further expect that the oldest ghosts are swept into
clusters of galaxies by the flow of structure formation. But
buoyancy of the probably very light radio plasma can produce some
relative motion between ghosts and the IGM gas. That buoyancy is
operating on radio plasma is impressively demonstrated by a recent
simultion of a buoyant rising radio blob in a cluster center,
which reproduces radio and X-ray observations of M87 strikingly
well \cite{churazov200sub}.

This might allow the ghosts in clusters or filaments of galaxies to
ascend to larger radii, until they get stopped by freely infalling
matter at the accretion shock.  It depends crucially on the topology
of this accretion shock surface if the ghosts are able to escape from
gravitationally bound structures as clusters and filaments of
galaxies, or not. It is therefore difficult to predict the spatial
distribution of ghosts. In this work we consider cases where the
ghosts are distributed as the galaxies, where they are more, and where
they are less clustered.

There are several observations supporting their existence and future
experiments might be able to see further signatures of ghosts.
\begin{itemize}
\item
The progenitor of radio ghosts, radio cocoons of active radio
galaxies, are observed and well studied. Since the sharp boundaries of
radio cocoons indicate that these objects are able to resists their
very turbulent birth, it is reasonable to assume that they also
survive the much weaker ambient turbulence.
\item
A radio ghost can be revived to radio emission by the compression in
a shock wave. This should lead to regions of diffuse synchrotron
emission within clusters of galaxies, where accretion or merger shock
waves are expected. Such objects are indeed observed, and are named
cluster radio relics. Models for such processes can be found in
En{\ss}lin et al. \cite*{1998AA...332..395E} and En{\ss}lin \& Krishna
(2000)\nocite{ensslin2000b}.
\item
Another class of diffuse radio emitter in galaxy cluster might also be
connected to ghosts: cluster radio halos. There is the speculative
possibility that the short-lived radiating electrons are replenished
by the decay of charged pions produced in hadronic collisions of
long-lived cosmic ray protons, which were released from their
confinement in radio ghosts \cite{Ringberg99}.
\item
Although at low energies, the electron population within ghosts should
remain mostly relativistic. Comptonization of the CMB photons by this
electron population should produce CMB distortions, which have a
spectral signature comparable to that of a grey absorber. This
signature is weak with an optical depth of the order of $10^{-7}$
\cite{ensslin2000a},
but principally observable. The energetically up-scattered CMB photons
might also be detectable.
\end{itemize}
For this work, to estimate the deflection of UHECR by ghosts, most of
their above discussed properties are not of any importance. Only the
magnetic fields are required, and their existence is nearly guaranteed
by the existence of radio galaxies, and the extremely low magnetic
diffusivity in extragalactic plasmas. Therefore, in the context of
this work, the term `radio ghost' can be simply translated into
`remnant magnetic fields of a former radio cocoon'. Even if turbulence
is able to disrupt ghosts, the shear motions should increase the
magnetic energy content of the fields, leading to an enhanced
efficiency in UHECR deflection. We summaries, that the existence of
regions in the IGM filled with remnant magnetic fields of former radio
galaxies -- as it is assumed in this work -- is difficult to be
doubted.

\section{Properties of Radio Ghosts\label{sec:ghost prop}}
\subsection{The Distribution Function of Radio Ghosts}
The properties of a radio ghost depend on three factors: the state of
the radio plasma, as released by a radio galaxy (RG), the turbulent
history of the environment, and the present environment of the ghost.
We are mainly interested in ghosts located in filaments of galaxies,
where turbulence is less strong compared to galaxy clusters. In our
simplistic approach, we neglect therefore any environmentally induced
changes in the ghost properties. We assume that a ghost was originally
produced with a total energy $E_\gh$, and this energy did not change
significantly over cosmological times. The ghost consists of magnetic
fields and relativistic particles (electrons and protons), which
should be roughly in energy equipartition. The ghost's internal energy
density
\begin{equation}
  \label{eq:gh.equipart}
  \eps_\gh = \eps_B + \eps_\CR \approx 2\,\eps_B
\end{equation}
is related to it's internal pressure via the relativistic equation of
state, which the CR population and the tangled magnetic fields do follow:
\begin{equation}
  \label{eq:rel.eq.state}
  P_\gh = \frac{1}{3} \eps_\gh \approx \frac{2}{3} \eps_B\,.
\end{equation}
Since the ghost is in pressure equilibrium with its environment $P_\gh
= P$, the energy released by the RG can be written as the sum of the
internal energy of the ghost and the volume work required to produce a
cavity in the IGM for the ghost:
\begin{equation}
  \label{eq:gh.energy}
  E_\gh = \eps_\gh \, V_\gh + P\, V = 4\, P \, V_\gh\, .
\end{equation}
The relevant properties of a ghost, as volume $V_\gh$ and magnetic field
strength $B_\gh $, are therefore determined by $P$ and $E_\gh $:
\begin{equation}
  \label{eq:V}
  V_\gh(E_\gh, P)  = \frac{E_\gh}{4\, P}
\end{equation}
\begin{equation}
  \label{eq:B}
  B_\gh(P) = \sqrt{12 \,\pi\, P}
\end{equation}
Thus we need a model for the IGM pressure. The enthalpy of the gas is
increased during structure formation by compression of the flow, which
is driven by the evolving gravitational
potential\footnote{During structure formation the gas follows the energy equation
\begin{equation}
\frac{d}{dt} \left(\frac{1}{2} v_\gas^2  + \frac{\gamma}{\gamma-1
}\,\frac{P_\gas}{ \varrho_\gas} + \Phi_{\rm grav}\right) = 
 \frac{\partial \Phi_{\rm grav}}{\partial t},
\end{equation}
if heating and cooling processes can be neglected.  This equation can
be integrated, if we approximate the potential to be time independent,
yielding
\begin{equation}
\frac{1}{2} v_\gas^2  + \frac{\gamma}{\gamma-1}\,\frac{P_\gas}{
\varrho_\gas} = - (\Phi_{\rm grav} - \Phi_{\rm grav,o}). 
\end{equation}
Whenever the kinetic energy is small a unique relation between the potential
and the specific enthalpy is given. 
} . This picture neglects
non-gravitational heating and cooling processes. Cooling is only
important in the center of dense clusters of galaxies.  Whenever the
gas is subsonic, the pressure is related to the gas density via
\begin{equation}
  \label{eq:bias1}
  P(n_\gas) = P_{\rm o} \,(n_\gas/n_{\rm gas,o})^{\gamma}\,,
\end{equation}
where $\gamma = 5/3$ is the adiabatic index of the gas, otherwise the
pressure is lower. We neglect deviations from this equations assuming
that only a small fraction of gas has a sonic or higher velocity. This
is justified, since supersonic gas becomes soon shocked when it hits
the next accretion shock around a cluster or filament of galaxies.

Unfortunately, the distribution of gas in the local Universe is not
observed with current instruments. But the number density $n_\gal$ of
the galaxy distribution can be used to give a rough idea of the gas
and pressure distribution. We assume that the baryon fraction of the
matter density is approximately the same in all structures
($\varrho_\gas/\varrho_{\rm matter}= {\rm const}$), and that the bias
between optical galaxies and matter can be approximated by
\begin{equation}
n_\gal = n_{\rm gal, o} (n_\gas/n_{\rm \gas,o})^{b_\gal}.
\end{equation}
The galaxy bias can be found to be $b_\gal = 1.6$ by linearizing this
expression and comparison with the observed linear bias for optical
galaxies \cite{1999coph.book.....P}.
Thus we get a rough proportionality between galaxy density and
pressure:
\begin{equation}
\label{eq:P(ngal)}
  P(n_\gal) = P_{\rm o} \,(n_\gal/n_{\rm gal,o})^{\gamma/b_\gal}\sim
  n_\gal^{1.042}\,.
\end{equation}
This relation can be normalized by using the observed properties of
clusters of galaxies. A typical density and temperature of a cluster is
$n_e = 10^{-3} \, {\rm cm^{-3}}$ and $kT = 5$ keV, leading to $P_o =
P_{\rm cluster} = 10 \,{\rm keV\,cm^{-3}}$. We therefore get a ghost
magnetic field strength of $25\,\mu$G inside of galaxy clusters and
$0.54\,\mu$G in galaxy filaments (assuming $n_e = 10^{-5}\,{\rm
cm^{-3}}$).

The distribution of ghost energies is assumed to follow a Schechter
distribution
\begin{equation}
  \label{eq:press.schechter}
  n_\gh(E_\gh) \,dE_\gh\, = N_\gh\,
  \left( \frac{E_\gh}{E_*} \right)^{-\alpha_\gh}\,
  \exp \left( - \frac{E_\gh}{E_*} \right)
  \,\frac{dE_\gh}{E_*}\,,
\end{equation}
for $E_\gh > E_{\rm min}$, $n_\gh(E_\gh) = 0$ otherwise. Both the
normalization factor $N_\gh$ and the upper cutoff $E_*$ are allowed to
depend on the position, parameterized by the galaxy density
($N_\gh = N_\gh(n_\gal)$ and $E_*= E_*(n_\gal)$).

It is convenient to introduce the ghost-volume-filling factor
\begin{equation}
  \label{eq:Phi}
  \Phi_\gh = \int_{E_{\rm min}}^{\infty} \!\!\!\!\!\!\!\!\! dE_\gh
  V_\gh(E_\gh, P)\, n_\gh(E_\gh, n_\gal) ,
\end{equation}
and the average energy density in ghosts
\begin{equation}
  \label{eq:av ghost energy}
\epsilon_\gh = \frac{4}{3}\, \eps_\gh \,\Phi_\gh .
\end{equation}
The factor $\frac{4}{3}$ appears here, since our definition of the
energy of ghosts included the volume work done by the expanding
radio lobe. This part of the ghosts energy, which is $\frac{1}{4}$
of its total energy budget, is located in the IGM in the form of
heat, kinetic, and gravitational energy.

The normalization factor can therefore be expressed as
\begin{equation}
  N_\gh = \frac{\epsilon_\gh}{E_* \Gamma(2-\alpha_\gh, E_{\rm min}/E_*)}\,,
\end{equation}
where $\Gamma$ is the incomplete Gamma-function:
\begin{equation}
\Gamma(a,x) := \int_x^\infty \!\!\!\!\!\! dt\; t^{a-1}\,e^{-t}\;.
\end{equation}
The total number density in ghosts reads
\begin{equation}
n_\gh = \int_{E_{\rm min}}^\infty E_\gh\;
n_\gh(E_\gh) = \frac{\epsilon_\gh}{E_*}\frac{ \Gamma(1-\alpha_\gh,
E_{\rm min}/E_*)}{\Gamma(2-\alpha_\gh, E_{\rm min}/E_*)}\,.
\end{equation}

Finally we assume that the energy density in ghosts scales with the
galaxy density to some power $b_\gh$
\begin{equation}
\label{eq:epsilongh(ngal)}
 \epsilon_\gh (n_\gal) = \epsilon_{\gh \rm ,o} (n_\gal/n_{\rm \gal
 ,o})^{b_\gh}.
\end{equation}
$b_\gh = 1$ would correspond to constant energy output per
galaxy. This might be unrealistic, since the older, and bigger
galaxies in the denser regions are expected to have produced more
radio plasma during their lifetime. We therefore also adopt $b_\gh = 2$
in order to have a ghost distribution more clustered than the galaxies.

On the other hand, buoyant motion of the ghost might have allowed a
significant fraction to escape from clusters and filaments of
galaxies. This could have let to a distribution of ghosts which is
less clustered than the galaxies. We mimic this case by also using
$b_\gh = 1/2$.

\subsection{Guessing the Parameters}
The parameters, which are still undetermined, are $\epsilon_{\gh,\rm
o}$, $\alpha_\gh$, $E_{\rm min}$, and $E_*(n_\gal)$.

The progenitors of ghosts are RGs. We therefore try to use the
observed radio luminosity function (RLF) of radio galaxies to get an
idea about the shape and normalization of the ghost size
distribution. The present day spatially averaged number density of
ghosts is simply given by the time integral of their birth rate
\begin{equation}
\label{eq:integral}
\bar{n}_\gh(E_\gh) = \int \, dt \, \dot{\bar{n}}_\gh(E_\gh, t).
\end{equation}
The birth rate is given by the number density of visible radio
galaxies divided by their typical lifetime $t_{\rm visible}$. We
assume that the time the galaxy is active $t_{\rm active}$ and the
time the galaxy is visible $t_{\rm visible}$ are the same for all
galaxies ($t_{\rm active} = t_{\rm visible} = t_\rg = 3\cdot 10^7$
yr).  This is a typical value out of the observed range of ages of
large expanded radio lobes, which are of the order of 100 Myr
\cite{1987MNRAS.225....1A,1994A&A...285...27K,1998MNRAS.298.1113V,1998MNRAS.297L..86S}.Although
these fairly robust age estimates are based on interpreting the
observed radio spectral steepening in terms of the inverse Compton
losses of the relativisitic electrons against the ubiquitous
cosmic microwave background photons uncertainties and scatter are
likely up to one order of magnitude. We note, that the assumed age
only affects the distribution of the total ghost energy budget
onto small and large ghosts sizes, but not the budget itself.

We further assume that the jetpower $q_\jet$ and radio
luminosity $L_\nu$ of a RG does not change during its active
phase. These are very coarse assumptions. But we hope they are
justified since we use them only to get a feeling for the ghosts size
distribution.  Radio galaxies with jetpower $q_\jet$ produce therefore
ghosts with the energy
\begin{equation}
E_\gh = \frac{1}{2}\, q_\jet\, t_\rg
\end{equation}
with a birth rate of
\begin{equation}
\label{eq:ghost.birth}
\dot{\bar{n}}_\gh(E_\gh, t) = 2 \, \frac{n_\jet(q_\jet)}{t_\rg} \, \frac{d
q_\jet}{d E_\gh} = 4\, \frac{n_\jet(q_\jet, t)}{t_\rg^2} ,
\end{equation}
where $n_\jet(q_\jet, t)\, dq_\jet$ is the jetpower distribution
function of RGs. The number two appears since RG have typically two
radio lobes.

The jetpower distribution function can be derived from the radio
luminosity function of RG with the additional assumption that there is
a unique relation between radio luminosity and jetpower. En{\ss}lin et
al. (1997)\nocite{1997ApJ...477..560E} have fit a power-law relation
to the jetpower derived by Rawlings and Saunders
(1991)\nocite{1991Natur.349..138R} of a sample of radio
galaxies. Their values are based on minimum energy arguments and age
estimates. The real energy of the radio plasma can easily be much
higher than the minimal energy estimate by some factor $f_\power >1 $
due to the presence of relativistic protons, low energy electrons or
deviations from equipartition between particle and field energy
densities. A rough estimate of $f_\power$ can be derived from
observations of radio lobes embedded in the intra-cluster medium (ICM)
of clusters of galaxies. They show a discrepancy of the thermal
ICM-pressure to the pressure in the radio plasma following minimal
energy arguments by a factor of $5 - 10$, even if projection effects
are taken into account \cite{1992A&A...265....9F}.
Since also a filling factor smaller than unity of the radio plasma in
the radio lobes can mimic a higher energy density we chose $f_\power =
3$ in order to be conservative.

The jetpower-radio luminosity correlation at $\nu = 2.7 $ GHz then
reads
\begin{equation}
\label{eq:jetpower}
q_{\rm jet} = a_\nu  \,(L_{\nu}/({\rm Watt\,
Hz^{-1}}\,h_{50}^{-2}))^{b_\nu}\, f_\power
\end{equation}
for which $b_\nu =0.82 \pm 0.07$, $a_\nu =10^{45.28 - 26.22
\cdot b_\nu \pm 0.18}\,{\rm erg\, s^{-1}}\, h_{50}^{-2}$
\cite{1997ApJ...477..560E}.
%
This relation allows us to translate the
observed radio luminosity function into a jetpower distribution
function. We adopt different radio-luminosity functions parameterized
by Dunlop \& Peacock \cite*{1990MNRAS.247...19D} and integrate
Eq. \ref{eq:integral} using Eqs. \ref{eq:ghost.birth} and \ref{eq:jetpower} in an
Einstein de Sitter Cosmology up to redshift $z = 10$ (the result
depends only weakly on the upper cutoff due to the strong decline of
the used RLFs beyond $z=3$, the pure luminosity evolution (PLE) RLF is
only integrated up to $z=4$). We further calculate the ghost
distribution function for $b_\nu = 0.7, 0.82, 1$ in order to show the
dependence on the uncertainties. The results can be seen in
Fig. \ref{fig1}.

Fig. \ref{fig1} shows that $\alpha_\gh \approx 2$, which we
adopt in the following. A reasonable lower cutoff is then $E_{\rm min}
= 6\cdot10^{57}$ erg (corresponding to $L_\nu = 10^{23}\,{\rm Watt\,
Hz^{-1}}$). The ghost distribution does not need to have a real cutoff
there, but the part which is accessible to radio observations of the
parent radio galaxies ends there. In order to conservative we do not
extrapolate to lower energies.

The radio luminosity function itself becomes dominated by starburst
galaxies at smaller radio luminosities. This gives a second contribution
to the magnetization of the IGM from magnetized galactic winds. An
extrapolation of the strongly redshift dependent, recent starburst
history of galaxies to high redshifts ($z=10$) indicates that a
substantial cosmological volume might be magnetized by such winds
\cite{1999ApJ...511...56K}.
However, the fields resulting from this process are expected to be
tangled on small, galactic scales, and therefore less effective
scattering centers for UHECR as radio ghosts.

\begin{table}
\begin{tabular}{clcc}
RLF & $b_\nu$ & $\bar{\epsilon}_\gh /(10^{66}\,{\rm erg\, Gpc^{-3}})$
& $\bar{n}_\gh /( 10^8\, {\rm Gpc}^{-3})$\\
\hline
PLE  & 0.7  & 2.80 & 0.28\\
PLE  & 0.82 & 2.19 & 0.28\\
PLE  & 1.0  & 1.91 & 0.28\\
RLF2 & 0.7  & 5.05 & 1.11\\
RLF2 & 0.82 & 3.32 & 1.11\\
RLF2 & 1.0  & 2.55 & 1.11\\
\end{tabular}
\caption[]{\label{tab:totaljetpower} Cosmological energy output of
radio galaxies for two radio luminosity functions of Dunlop \& Peacock
\cite*{1990MNRAS.247...19D} which are used in
Fig. \ref{fig1}. The last column gives the total number
density of ghosts in the ernergy range plotted in
Fig. \ref{fig1}.}
\end{table}

An integration of the ghost energy distribution functions shown in
Fig. \ref{fig1} gives the total energy output of radio
galaxies per volume. The results do not depend on the assumed lifetime
$t_\rg$ of radio galaxies, they scale linearly with $f_\power$, and
depend only moderately on $b_\nu$. Detailed results are given in
Tab. \ref{tab:totaljetpower}. The energy input in form of radio plasma
is roughly $\bar{\epsilon}_\gh = 10^{66...67} \,{\rm erg\, Gpc^{-3}}$,
it is distributed over $\bar{n}_\gh = 10^{7...8}$ ghosts per Gpc$^3$,
which have an average energy of $\bar{\eps}_\gh = 10^{58 ... 59}$
erg. Note that the X-ray background, which is believed to be dominated
by AGN emission, corresponds to an injection energy of $\approx 3
\cdot 10^{67} \,{\rm erg\, Gpc^{-3}}$ (Chokshi \& Turner 1992; Soltan
1982)\nocite{1992MNRAS.259..421C,1982MNRAS.200..115S}. Either the
X-ray energy losses of AGNs exceed the radio plasma release, or the
jetpower is strongly underestimated here and $f_\power = 10 ... 30$
would be more realistic. Even such a strong energy input into the IGM
seems to be consistent with present the day limit to the
Comptonizarion of the IGM
\cite{ensslin2000a,1998AA...333L..47E}.

The next parameter we have to fix for our model is the maximal ghost
energy.  $E_*(n_\gal)$ can be roughly estimated from observations as
shown in the following.  The progenitor of large ghosts in rare
regions should be giant radio galaxies (GRGs). Their linear size
distribution seems to have a sharp cutoff at 3 Mpc $h_{50}^{-1}$
(Ishwara-Chandra \& Saikia 1999)\nocite{1999MNRAS.309..100I}. There is
only one GRG significantly exceeding this: 3C236 has a linear size of
5.7 Mpc $h_{50}^{-1}$. This could be -- for example -- due to a very
low pressure in the environment of 3C236. For a typical galaxy
filament environment, we assume the linear size of GRG to have a
cutoff at 3 Mpc $h_{50}^{-1}$.  The diameter of one of the two radio
lobes is roughly $\frac{1}{3}$ of the linear size of the whole GRG. We
adopt therefore a maximal radius of our (spherically approximated)
ghosts in the dilute environment of galaxy filaments of $r_{*\rm} =
0.5$ Mpc $h_{50}^{-1}$. The distribution function of ghost sizes
\begin{equation}
  \label{eq:size-distr} f_{\rm o}(r)\, dr =
  \frac{3\,(r/r_{*})^{2-3\alpha_\gh}}{\Gamma(1-\alpha_\gh, E_{\rm
  min}/E_*)}\, \exp \left( - \frac{r^3}{r_{*}^3} \right)
  \frac{dr}{r_{*}}
\end{equation}
has indeed a very sharp cutoff at $r_{*}$, as required by the
observation of GRG. This $r_{*}$ is related to $E_*$ via Eqs.
\ref{eq:gh.energy} and \ref{eq:rel.eq.state}:
\begin{equation}
  \label{eq:r*} E_* = \frac{16}{3} \,\pi\, r_{*}^3 P \approx
  \frac{32}{9} \,\pi\, r_{*}^3\, \eps_B = \frac{4}{9}\, r_{*}^3\,
  B_{\gh}^2
\end{equation}
Equipartition (or minimum energy) magnetic field strength of GRG are
of the order of a few $\mu$G (Ishwara-Chandra \& Saikia
1999)\nocite{1999MNRAS.309..100I}, so that we adopt a conservative
value of $B_{\gh} = 1 \mu$G for ghosts located in galaxy
filaments. From this we find that $E_* = 1.6 \cdot 10^{60}\, {\rm
erg}$, and that a corresponding RG had ejected twice as much energy
($3.2 \cdot 10^{60}\, {\rm erg}$), which seems to be
moderate. Comparison with Fig. \ref{fig1} shows that ghosts
with $10-100$ times higher energy are expected from the radio counts
(if $t_\rg$ is as big as $3\cdot 10^{7}$ yr). Such ghosts are also
required since a $10^{60}$ erg ghosts in a dense cluster environment
($n_{\rm e} = 10^{-3}\,{\rm cm^{-3}} \rightarrow B_\gh = 25 \mu$G) has
a diameter of $0.1$ Mpc, which is clearly below the observed maximal
size of radio lobes inside clusters. In order to allow a diameter of
$0.5$ Mpc inside a cluster one needs $E_* \approx 10^{62}$ erg. We
therefore assume that the maximal energy of a ghost scales linearly
with the gas density:
\begin{equation}
\label{eq:E*(ngal)}
E_* = 10^{62} \,{\rm erg} \, \frac{n_{\rm e}}{10^{-3}\,\rm cm^{-3}} =
10^{62} \,{\rm erg} \, \left( \frac{n_\gal}{n_{\gal \rm ,o}}
\right)^{1/b_\gal} ,
\end{equation}
where $n_{\gal \rm , o}$ is the galaxy density in the center of a
typical cluster of galaxies. Such a scaling might be justified due to
the fact that the most massive galaxies, which probably host the most
powerful AGNs, are the central elliptical galaxies within clusters.

For the purpose of our simulation it also makes sense to define the
average total cross section for the assemble of ghosts:
\begin{equation}
\bar{\sigma}_\gh = \int dr\; f(r)\, \pi\, r^2 = \pi\,r_*^2\;
\frac{ \Gamma(\frac{5}{3}-\alpha_\gh, E_{\rm
min}/E_*)}{\Gamma(1-\alpha_\gh, E_{\rm min}/E_*)}
\end{equation}

\subsection{Constructing the Ghost Distribution}

The scaling relations in Eqs. \ref{eq:P(ngal)},
\ref{eq:epsilongh(ngal)}, and \ref{eq:E*(ngal)} are normalized for a
reference density equal to the central galaxy density of a galaxy
cluster. We estimate $n_{\gal \rm ,o}$ by identifying the peaks in our
galaxy distribution. The energy density of ghosts at our reference gas
density $n_{\gal \rm ,o}$ (inside clusters) is given by
\begin{equation}
\epsilon_{\gh \rm ,o} = \frac{\bar{\epsilon}_\gh \,V}{ \int\, dV
\,(n_{\gal}/n_{\gal \rm ,o})^{b_\gh}} ,
\end{equation}
where $\bar{\epsilon}_\gh = 10^{66...67} \,{\rm erg\, Gpc^{-3}}$ and V
is the volume of our galaxy catalog. Whenever we use the CfA redshift
catalog we exclude volumes close to the galactic plane in our
estimate of V and the above integral due to the under-sampling of
galaxies there.

If the ghosts would get disrupted and mix with the ambient gaseous
medium a significant magnetization of the IGM would result.
Assuming complete mixing, we estimated the resulting magnetic
field strengh as a function of the local cosmic matter density in
a few scenarios ($\bar{\epsilon}_\gh = 10^{66...67} \,{\rm erg\,
Gpc^{-3}}$, $b_{\rm gh} = 0.5$ and $b_{\rm gh} = 1$). The field
strength as a function of density is plotted in Fig. \ref{fig12},
and the volume fraction of the different density regimes in Fig.
\ref{fig13}. Such magnetic fields are not excluded by Faraday
rotation measurements, which give limits of $B_{\rm IGMF}<nG$ for
a homogeneous background field with Mpc coherence-scales
\cite{1994RPPh...57..325K}, or $B_{\rm IGMF}<\mu$G if cosmological
fields are concentrated in the large scale structures like sheets
and filaments of galaxies
\cite{1998A&A...335...19R,1999ApJ...514L..79B}. The implied values
for cluster of galaxies of $B_{\rm ICM} \sim 0.3 ... 3 \mu$G are
also consistent with Faraday rotation measurements of magnetic
fields. Note, that shear flows due to cosmic structure formation
can significantly increase the strength of IGM magnetic fields
\cite{1999A&A...348..351D,1999ApJ...518..594R}.

\subsection{The Radio Ghosts Produced by Our Galaxy\label{milky way}}

The formation of the $M_{\rm bh} = 2.6\cdot 10^{6}\,M_\odot$ black
hole in the center of our galaxy (Eckart et al. 1997) was probably
accompanied by dissipative processes as radiation and ejection of
relativistic plasma in a temporary AGN. We assume that a fraction
$\epsilon_{\rm diss} = 0.1$ of the rest mass energy of the accreted
matter was dissipated and that $\epsilon_{\rm plasma}= 0.5$ of this
energy went into the relativistic plasma.
The radio ghosts formed by the remnant of this radio plasma got
therefore an energy budget of
\begin{equation}
E_{\rm gh} = \frac{\epsilon_{\rm plasma} \, \epsilon_{\rm diss}}{1 -
\epsilon_{\rm diss}} \, M_{\rm bh}\, c^2 = 2.6 \cdot 10^{59}\, {\rm
erg} \,.
\end{equation}
The term ${1 -\epsilon_{\rm diss}}$ in the denominator corrects for
the mass discrepancy between accreted mass and final black hole mass
due to dissipation of energy.  Eqs. \ref{eq:gh.equipart},
\ref{eq:rel.eq.state}, and \ref{eq:gh.energy} imply that the filled
volume is given by
\begin{equation}
V_\gh = \frac{3 \, E_{\rm gh}}{8\, \eps_B} = 0.083\, {\rm Mpc^3}
\left(\frac{B}{\mu{\rm G}} \right)^{-2}\,.
\end{equation}
Comparing this to the gyroradius $r_{\rm g}$ of a $10^{20}\, {\rm eV}$
cosmic ray proton shows the possible importance of the ghosts of our
own galaxy: $V_\gh \approx 66\,  r_{\rm g}^3 \,(B/\mu {\rm G})$.

But it cannot be assumed that the ghosts produced by our own galaxy
are located in the direct vicinity of our galaxy since the observed
peculiar velocity of 162 km/s with respect to our own galaxy group
(Rauzy \& Gurzadyan 1998) indicates differential motion between the
Milky Way and the local IGM. Radio ghosts produced by other members of
the local group may be nearby today, but this is already included into
our statistical approach described in Sec. \ref{sec:ghost prop}, where
the density of ghosts scales with the number density of galaxies.

\section{UHECR Flux at Earth}
\subsection{Scattering by Radio Ghosts}

Equation (5) can be used to calculate the intensity of the
magnetic field inside ghosts. The estimation of the topology
of the internal field is, however, a much more difficult matter.

The generation of any ghost was a turbulent event fed by the parent
radio galaxy. This process certainly produced vortices at all
scales. Ghosts, however, have been dynamically decoupled afterwards
and, therefore, smaller scale wiggles should have been dissipated
over time. Hence, most of the remaining power in the field probably
lays at large scales at present.

In Fig. \ref{fig2} we exemplify one possible internal magnetic
field topology. The shown magnetic field has null perpendicular
component at the surface and is therefore decoupled from the
environment field. The amplitude of the fluctuations obeys a power
law power spectrum of spectral index $\xi = 4$.

In order to analyze the scattering capability of ghosts and to
estimate how critical is the uncertainty in the actual topology of
the field to the present study, we performed numerical simulations
using spherical radio ghosts, similar to the one in Fig.
\ref{fig2}. Different power spectral indexes $\xi$ and radius were
considered. We illuminate the ghosts with a beam of test
particles, protons, with a power law energy spectrum, $dN/dE
\propto E^{-3}$, and energy $E > 4 \times 10^{19}$ eV. The angular
distribution of the scattered particles is shown in Fig. 3.a for
ghosts of different sizes and $\xi=5/3$, $3$ and $4$. The radii of
the ghosts are given in terms of the average gyroradius of the
injected UHECR spectrum. As a comparison, an isotropic scattering
particle distribution is also shown in the same figure. It can be
seen that ghosts are very efficient at scattering UHECR regardless
of power spectral index. Forward scattering becomes dominant only
at the low end of the ghost size distribution, when the ghost's
radius becomes comparable to the average gyroradius of the
particles. Figure 3.b is the same as figure 3.a but for a
monoenergetic proton injection at $E = 10^{20}$ eV. The same
ghosts radii as in fig. 3.a (physical units) were used, and so
they are down by a factor of two in units of UHECR gyroradius in
fig. 3.b. The stronger dominance of forward scattering for the
smallest ghosts can be appreciated, as well as for the flattest
fluctuation spectrum ($\xi = 5/3$). Nevertheless, as previously
argued, small scale magnetic turbulence inside a typical ghost
should have been dissipated over time, and a steep spectrum of
magnetic fluctuations is expected.

Therefore, given the current uncertainties in radio ghost shape, size,
magnetic field topology and spatial distribution, isotropic
scattering is, very likely, an acceptable assumption and it will
be used in what follows.

\subsection{Numerical Model}

We use Monte Carlo numerical simulations to track UHECR propagation
through the intergalactic medium and to evaluate their arrival
distribution at Earth.

We start from the basic assumption that most UHECR are protons
of extragalactic origin whose sources aggregate spatially as
either luminous or cold DM.

Charged particles, even at the extreme energies considered,
are coupled to the intervening magnetic fields. From this
point of view, the propagation region can be divided in
several components: (a) the sources and their immediate
neighborhood, (b) the intergalactic medium, (c) the galactic
halo, (d) the galactic disk, (e) the heliosphere and (f) the
magnetosphere.

The close environment of the sources can be included
inside the sources just by re-defining them as large as needed.
This should be acceptable in general since, the newly defined
source, would still be below the resolution capability of any
current experiment.

The heliospheric and magnetospheric fields, despite their
relatively large values, are confined to small volumes and,
from the point of view of deflection, are negligible.

The galactic disk magnetic field is one to two orders of
magnitude smaller than the latter fields, but fills
regions at least 8 orders of magnitude larger. However,
exception made of lines of sight crossing directly the
galactic bulge or a relatively narrow strip surrounding the
galactic plane, the disk field has probably little impact
on UHECR deflection.

Consequently, the two main factors responsible for blurring
pointing information, are the intergalactic (IGMF) and the
galactic halo magnetic field.

Few is known regarding the large scale structure of the IGMF and only
sparce rotation measurements are available to constrain its intensity
\cite{1994RPPh...57..325K,Vallee97}.  Basically, two extreme
scenarios can be imagined that satisfy the observational
constraints. In one of them, the field is compressed inside walls and
filaments, where it attains high intensity ($\sim 0.1 - 1 \mu$G) and
high degree of correlation (over scales of up to tens of Mpc), leaving
unmagnetized voids behind \cite{Ryu98}.  We will call this the
laminar IGMF model.  A second possibility is that of a more evenly
distributed field, whose intensity scales somehow with matter and is
coherent on scales $L_{c} \approx {\rm 1\, Mpc}\, (B/{\rm
nG})^{-2}$. A numerical model for the latter case is an ensemble of
cells of characteristic size $L_{c}$, with uniform field inside but
randomly oriented with respect to adjacent cells. The galaxy density
is used to scale $B$ and $L_{c}$. We call this model the cellular
IGMF.

The first IGMF scenario can affect propagation in a radical way,
with UHECR mainly constrained to the interior of large scale
structures and drifting along walls and filaments. The
observable flux depends critically on the location of the
observer and on the details of the actual topology of the field
inside the GZK-sphere. Isotropy can be attained in a rather
natural way, but the results are strongly model dependent.
The propagation of UHECR in this IGMF has been treated
in an earlier work \cite{GMT_APJL98_IGMF}, and will not
be considered here.

A similar uncertainty exists around the topology of the magnetic field
inside the galactic halo. Nevertheless, unless there is a large scale
magnetized galactic wind that structures this field and maintains a
large azimuthal component up to large galactocentric distances, the
halo should be unimportant regarding isotropization. If a magnetized
galactic wind does exist on the scale of several hundreds of kpc, then
it would be the dominant factor in the determination of the UHECR flux
at Earth \cite{Ahn_etal_GW00,Biermann_etal_Erice99}; therefore,
extragalactic propagation details would be mostly irrelevant.

Therefore, we restrict the present analysis to a cellular
IGMF and neglect the galactic halo.

The same procedure as in
\cite{GMT_Durban_IGMF97,GMT_APJL98_cluster} is used in
the description of the cell-like spatial structure of the
IGMF. The cell size is given by the correlation length,
$L_{c} \propto B_{IGMF}^{-2}(r)$. The intensity of the
IGMF, in turn, scales with gas density as
$B_{IGMF} \propto n_{gas}^{\eta}(r)$ and the proposed IGMF
value at the Virgo cluster ($\sim 10^{-7}$ G, \cite{Arp88})
is used as the normalization condition. We use $\eta = 0.5$,
as a compromise between a frozen-in field ($\eta = 2/3$)
and Valle\'e's (1997) estimate ($\eta \sim 0.35$), which
may be too flat due to the assumed values for the magnetic
field in superclusters and larger scales. Nevertheless,
tests have been conducted for different values of
$\eta$ covering the previous interval, and the scaling is
not critical to our conclusions.

The formulation of section 2 is used to define the size
and spatial distribution of radio ghosts. This requires the
knowledge of the gas density distribution inside the
simulation volume. Actually, as galaxies are easier
to survey over {\bf large} volumes, equation(9) is used
to transform between galaxy and gas density distributions.
There are, however, serious an unavoidable bias and
sampling problems inherent to galaxy surveys. Therefore,
we relay on both galaxy surveys and cold DM large scale
structure simulations to perform independent evaluations.

The sources of UHECR are distributed according to either
the galaxy or cold DM distributions respectively.

The 1999 version of the CfA catalog \cite{Huchra_ZCAT} is used
to characterize the galaxy distribution.

Cold DM simulation data are from Springel et al. (in preparation).
They carried out simulations that mimic the Local Universe. The
initial conditions of these simulations have been constrained by the
redshift survey of IRAS galaxies. As a result, the simulations develop
the same local large-scale structure (e.g., the Great Attractor and
Cetus Wall; clusters like Virgo and Coma are also found at the right
place).

The scenario is completed by the introduction of radio
ghosts according to the formulation of section 2.

Relativistic test  particles (UHECR protons) are injected at the sources
with a spectrum $dN/dE \propto E^{-2}$ and propagated through
the intergalactic magnetic field up to the detector on Earth.
Adiabatic energy losses due to redshift, pair production
and photo-pion production due to interactions with the
cosmic microwave background radiation (CMBR) are also
included \cite{Berezinsky88,Achterberg99}.

The flux at Earth can be divided into two components: (a) a
{\it direct \/} radiation field, constituted by particles
that fly from source to detector without encountering
ghosts and (b) a {\it diffuse \/} radiation field, comprising
particles which underwent at least one encounter with a
radio ghost.

\subsection{Different Scenarios}

Fig. \ref{fig4} is meant as a control. It shows the Aitoff
projection of the two-dimensional arrival probability density
(galactic coordinates with the anti-galactic center at the center
of the figure), for sources distributed according to nearby
luminous matter (CfA catalog,  \cite{Huchra_ZCAT}) inside $100$
Mpc and no radio ghosts. The same procedure as in
\cite{GMT_Durban_IGMF97,GMT_APJL98_cluster} is used in the
description of the intergalactic magnetic field (IGMF): a
cell-like spatial structure, with cell size given by the
correlation length, $L_{c} \propto B_{IGMF}^{-2}(r)$. The
intensity of the IGMF, in turn, scales with luminous matter
density, $\rho_{gal}$ as $B_{IGMF} \propto \rho_{gal}^{0.3}(r)$
\cite{Vallee97} and the observed IGMF value at the Virgo cluster
($\sim 10^{-7}$ G, \cite{Arp88}) is used as the normalization
condition. The mask covers the plane of the galaxy, where the
actual distribution of galaxies is not well known due to
obscuration by dust. The curved, thick line is the celestial
equator. Northern hemisphere is the sky patch to the right,
enclosed by that line.

Superimposed on the figure are the available events with
$E > 4 \times 10^{19}$ eV observed by AGASA
(47 events \cite{AG_1e19}),
Haverah Park (27 \cite{HP_4e19}), Yakutsk (24 \cite{YK_4e19})
and Volcano Ranch(6 \cite{VR_4e19}).

The arrival probability contours trace roughly the local
large scale structure. Distinguishable observational
signatures should be
expected towards the region of the Southern branch of the
supergalactic plane (to be observed in the near future by
the Auger experiment) at $l \sim 45^{o}$, the lines of sight
to the more distant Pisces-Perseus wall and Perseus cluster
and, very prominently, towards a large area surrounding the
Virgo and Ursa Major clusters.

The actually observed distribution of UHECR is clearly
much more isotropic than what one would expect under the
implicit assumptions in Fig. \ref{fig4}.

In figures 5, 6 and 7, we show how the UHECR arrival distribution
function would be modified by the presence of radio ghosts in
the intergalactic medium for different values of $b_{gh}$ and
$\alpha_{gh} = 2.0$. The results are insensitive to $\alpha_{gh}$
in the range of interest. The most important parameter is
$b_{gh}$, defined in equation (17), which tells how are ghosts
distributed with respect to galaxies. $b_{gh}=1$ implies the same
spatial distribution as galaxies and larger values a more clustered
distribution.

It can be seen from Fig. \ref{fig5} and \ref{fig6}, that clustered distributions
of ghosts produce no noticeable effects in the observed UHECR flux
at Earth.

In fact, if these angular distributions were observed with the same
exposure in declination as AGASA's after 7.5 years of integration, the
observed (ensemble averaged) amplitude of the first harmonic
\cite{Linsley_1st_harmonic} would vary only by $\sim 5\%$ when going
from Fig. \ref{fig4} to \ref{fig6}. This is basically because a distribution of
scatterers that is more clustered than the sources is, in practical
terms, equivalent to increase the physical size of the sources, which
is irrelevant from the point of view of all sky anisotropies.

The result is different when the scatterers are distributed in a
larger volume than the sources. This is the case in figure 7, where
$b_{gh}=0.5$. In this scenario it is assumed that e.g. radio ghosts
buoy out of cosmological structures, creating thick halos around walls
and filaments, permeating voids to some extent.  This diminishes
considerably the direct component (Fig. \ref{fig7}a) and accounts for
a considerable increase in the diffuse component (Fig. \ref{fig7}b)
which becomes, by far, dominant.  The composite flux
(Fig. \ref{fig7}c) still shows a smooth, large scale gradient towards
the region of the Virgo cluster but is much more isotropic that any of
the previous scenarios.

A more quantitative picture can be obtained from Fig. \ref{fig8}.
An ensemble of 1000 independent samples, of 50 UHECR each,
was built by observing the arrival distribution function
in Fig. \ref{fig7}c with an exposure equivalent to AGASA's.
The resultant distribution function of the amplitude of the
first harmonic is shown, in comparison with the actually
observed values by Yakutsk, AGASA, Haverah Park and Volcano
Ranch (see, \cite{GMT_AAW_DM_halo}). It can be seen
that $b_{gh}=0.5$ is the upper limit for the clustering
of ghosts that is compatible with the observed anisotropy.
However, it is also clear that that an even smoother
distribution of ghosts in intergalactic space would be much
more acceptable. Increasing the volume average ghost energy
density by a tolerable factor of 3 would help to move the
results in the right direction, but it wouldn't be a
determinant factor by itself.
Rephrased in other words, and looking at
Fig. \ref{fig9}, the mean-free path of UHECR for interactions with
ghosts in the intergalactic medium, should be reduced
to less than a few Mpc at any location in the propagation
region, i.e., walls and filaments as well as voids, to obtain
a level of isotropy consistent with observation.

Whether such a smooth spatial distribution of radio ghosts is
physically reasonable is a very complicated matter which we leave
open.  The possibility that radio ghosts are able to escape by
buoyancy from the denser environments in which they were produced
exists.  However, specific large scale structure hydrodynamic
simulations should be performed to decide whether buoyancy is actually
operational and, furthermore, a deeper knowledge of the composition
and mechanical properties of the radio plasma is necessary in order to
decide if ghosts are stable enough to survive as an entity under these
conditions.

As was mentioned previously, there are uncertainties associated
with unavoidable biases and sampling incompleteness associated
with galaxy surveys, as is the case with the CfA catalog used up
to here. This problem can be specially critical in this analysis,
since the results depend on an absolute normalization of the
density as a function of depth into the local universe. To check
the extension of the distortions occurring in our previous
analysis, we repeated our calculations using the distribution of
cold DM, calculated by large scale structure hydrodynamic
simulations (Springel, in preparation). They carried out
simulations that mimic the Local Universe, developing the observed
local large-scale structure.

Fig. \ref{fig10} and \ref{fig11} show the arrival flux for
$\alpha_{gh}=2.0$ and $b_{gh}=2.0$ and $0.5$ respectively for the
simulation using the CDM model. The resultant UHECR flux is plotted in
an Aitoff projection in supergalactic coordinates. The large spot near
the center of Fig. \ref{fig10} is the cosmic ray image from the Virgo
cluster. The central band is the supergalactic plane. Note that the
supergalactic plane is not such a neatly defined structure in UHECR
with average energy $E \stackrel{~}{>} 4 \times 10^{19}$ eV.  It is
clear that, regarding isotropization of the UHECR flux, the same kind
of effect is present in these new scenarios: the dominant signature
from Virgo at $b_{gh}>1$ is strongly diluted at $b_{gh} \le 0.5$.

\section{Discussion}

The, to some extent unexpected, isotropy in the UHECR flux
observed particularly by Yakutsk and AGASA, is a very
strong constraint to production and propagation models alike.
Most of the scenarios proposed in the literature should
produce a sizable anisotropy as either extragalactic luminous
or dark matter is normally associated with the invoqued
particle sources.
If that is really the case, then it is our understanding
of the topology and intensities of the intervening magnetic
fields that is critically incomplete. The problem could reside
inside the network of walls and filaments, in the interior of the
large surrounding voids, or even in our {\bf nearby} environment,
namely the Galaxy halo.

In the present work we explore the possibility that radio ghosts,
blobs of magnetized radio plasma remnant from past periods of
activity in radio galaxies, being able to scatter UHECR in the
intergalactic medium. Such a process could, in principle,
degrade the direct incoming flux from the sources and build
up a diffuse UHECR component large enough to be responsible
for the observed degree of isotropy.

Our results show that, over the most conservative region
of the radio ghost parameter space, such isotropization is not
possible. This stands not from an inability of ghosts to scatter
UHECR, but mainly from the fact that, under general conditions,
ghosts should tend to cluster more strongly than the sources of
the particles.

If, however, radio ghosts are able to buoy out into the
surroundings of the dense large scale structures and into voids,
while surviving the process, UHECR isotropy could be obtained
in those cases in which the mean free path for interactions with
ghosts is reduced below some few Mpc all over the propagation
region ($\sim b_{gh} < 0.5$).

Finally, one can question the basic assumption of this work, that
radio ghosts are able to survive for cosmological times as intact
objects. If radio ghosts mix with the ambient medium a significant
magnetization of the IGM should result, which can be sufficiently
strong in order to explain observed magnetic fields in clusters of
galaxies. These magnetic fields would have short coherence length, and
therefore do not violate Faraday rotation measurements. The ability of
the fields to deflect UHECR particles would be increased compared to
the case of intact, non-mixing ghosts for the following reasons: the
average deflection of particles in a ghosts vicinity is proportional
to $V_{\gh}\,B_{\gh}$ (cross section $V_\gh^{2/3}$ times scattering
angle, the latter is proportional to $B_\gh$ times $V_\gh^{1/3}$). For
a fixed magnetic energy budget $E_{B_\gh} = V_{\gh} \,B_\gh^2/(8\pi)$
of the ghosts, this implies that the scattering efficiency scales with
$V_{\rm gh}^{1/2}$ and therefore increases with increasing size.  For
fixed magnetic flux ($B_\gh \propto V^{-2/3}$) the scattering
efficiency inreases with the size as $V_{\rm gh}^{1/3}$.  This
demonstrates that relaxing the hypothesis of compact ghosts makes this
scenario even more powerful.\\[2em]

\noindent {\bf Acknowledgments}\\ \noindent The authors wish to
thank Dr. Simon White, Dr. Volker Springel, Dr. Avishai Dekel, Dr.
Gerard Lemson and others for kindly making available the DM
distribution from their large-scale-structure numerical simulation
of the local Universe using a constrained realization of the
initial perturbation field. It is a pleasure to acknowledge the
constructive comments of the anonymous referee. GMT is partially
supported by the Brazilian agencies FAPESP and CNPq. He thanks the
MPA for the hospitality during his visit. TAE was temporary
supported by the {National Science and Engineering Research
Council of Canada} (NSERC).

\bibliography{aamnem99,tae,gmt}

\begin{thebibliography}{}

\bibitem[\protect\astroncite{{Achterberg} et~al.}{1999}]{Achterberg99}
{Achterberg}, A., {Gallant}, Y.~A., {Norman}, C.~A., and {Melrose}, D.~B.,
  1999,
\newblock {MNRAS submitted, astro-ph/9907060}

\bibitem[\protect\astroncite{{Afanasiev}}{1995}]{YK_4e19}
{Afanasiev}, B. e.~a., 1995,
\newblock {Proceedings of 24th International Cosmic Ray Conference (Roma)} {2},
  796

\bibitem[\protect\astroncite{{Ahn} et~al.}{2000}]{Ahn_etal_GW00}
{Ahn}, E.-J., {Medina Tanco}, G.~A., {Biermann}, P.~L., and {Stanev}, T., 2000,
\newblock {astro-ph/9911123}

\bibitem[\protect\astroncite{{Alexander} and
  {Leahy}}{1987}]{1987MNRAS.225....1A}
{Alexander}, P. and {Leahy}, J.~P., 1987,
\newblock {\mnras} {225}, 1

\bibitem[\protect\astroncite{{Arp}}{1988}]{Arp88}
{Arp}, H., 1988,
\newblock {Phys. Lett. A} {129}, 135

\bibitem[\protect\astroncite{{Berezinsky} and
  {Grigor'eva}}{1988}]{Berezinsky88}
{Berezinsky}, V.~S. and {Grigor'eva}, S.~I., 1988,
\newblock {A \& A} {199}, 1

\bibitem[\protect\astroncite{{Bhattacharjee} and {Sigl}}{2000}]{Sigl_TD}
{Bhattacharjee}, P. and {Sigl}, G., 2000,
\newblock {Phys. Rept.} {327}, 109

\bibitem[\protect\astroncite{{Biermann} et~al.}{2000}]{Biermann_etal_Erice99}
{Biermann}, P.~L., {Ahn}, E.-J., {Medina Tanco}, G.~A., , and {Stanev}, T.,
  2000,
\newblock {Nucl. Phys. B} {87}, 417

\bibitem[\protect\astroncite{{Bird} et~al.}{1999}]{Bird98}
{Bird}, D.~J., {Dai}, H.~Y., {Dawson}, B.~R., {Elbert}, J.~W., {Huang}, M.~A.,
  {Kieda}, D.~B., {Ko}, S., {Loh}, E.~C., {Luo}, M., {Smith}, J.~D.,
  {Sokolsky}, P., {Sommers}, P., and {Thomas}, S.~B., 1999,
\newblock {\apj} {511}, 739

\bibitem[\protect\astroncite{{Blasi} et~al.}{1999}]{1999ApJ...514L..79B}
{Blasi}, P., {Burles}, S., and {Olinto}, A.~V., 1999,
\newblock {\apjl} {514}, L79

\bibitem[\protect\astroncite{{Chokshi} and
  {Turner}}{1992}]{1992MNRAS.259..421C}
{Chokshi}, A. and {Turner}, E.~L., 1992,
\newblock {\mnras} {259}, 421

\bibitem[\protect\astroncite{{Churazov} et~al.}{2000}]{churazov200sub}
{Churazov}, E., {Br{\"u}ggen}, M., {Kaiser}, C.~R., {B{\"o}hringer}, H., and
  {Forman}, W., 2000,
\newblock {\apj} submitted,
\newblock astro-ph/0008215

\bibitem[\protect\astroncite{{Cronin}}{1992}]{Cronin92}
{Cronin}, J.~W., 1992,
\newblock {Nucl. Phys. (Proc. Suppl.)} {28B}, 213

\bibitem[\protect\astroncite{{Dolag} et~al.}{1999}]{1999A&A...348..351D}
{Dolag}, K., {Bartelmann}, M., and {Lesch}, H., 1999,
\newblock {\aap} {348}, 351

\bibitem[\protect\astroncite{{Dubovsky} and {Tinyakov}}{1998}]{Dubovsky_DM}
{Dubovsky}, S. and {Tinyakov}, P., 1998,
\newblock {astro-ph/9802382}

\bibitem[\protect\astroncite{{Dunlop} and
  {Peacock}}{1990}]{1990MNRAS.247...19D}
{Dunlop}, J.~S. and {Peacock}, J.~A., 1990,
\newblock {\mnras} {247}, 19

\bibitem[\protect\astroncite{{En{\ss}lin}}{1999}]{Ringberg99}
{En{\ss}lin}, T.~A., 1999,
\newblock in P.~S. H.~B{\"o}hringer, L.~Feretti (ed.), {Ringberg Workshop on
  `Diffuse Thermal and Relativistic Plasma in Galaxy Clusters'}, Vol. 271 of
  {MPE Report}, p. 275,
\newblock astro-ph/9906212

\bibitem[\protect\astroncite{{En{\ss}lin} et~al.}{1998a}]{1998AA...332..395E}
{En{\ss}lin}, T.~A., {Biermann}, P.~L., {Klein}, U., and {Kohle}, S., 1998a,
\newblock {\aap} {332}, 395

\bibitem[\protect\astroncite{{En{\ss}lin} et~al.}{1997}]{1997ApJ...477..560E}
{En{\ss}lin}, T.~A., {Biermann}, P.~L., {Kronberg}, P.~P., and {Wu}, X.-P.,
  1997,
\newblock {\apj} {477}, 560

\bibitem[\protect\astroncite{{En{\ss}lin} and {Kaiser}}{2000}]{ensslin2000a}
{En{\ss}lin}, T.~A. and {Kaiser}, C.~R., 2000,
\newblock {\aap} {360}, 417

\bibitem[\protect\astroncite{{En{\ss}lin} and {Krishna}}{2000}]{ensslin2000b}
{En{\ss}lin}, T.~A. and {Krishna}, G., 2000,
\newblock {submitted to \aap}

\bibitem[\protect\astroncite{{En{\ss}lin} et~al.}{1998b}]{1998AA...333L..47E}
{En{\ss}lin}, T.~A., {Wang}, Y., {Nath}, B.~B., and {Biermann}, P.~L., 1998b,
\newblock {\aap} {333}, L47

\bibitem[\protect\astroncite{{Feretti} et~al.}{1992}]{1992A&A...265....9F}
{Feretti}, L., {Perola}, G.~C., and {Fanti}, R., 1992,
\newblock {\aap} {265}, 9

\bibitem[\protect\astroncite{{Greisen}}{1966}]{GZK_G66}
{Greisen}, K., 1966,
\newblock {Phys.Rev.Letters} {16}, 748

\bibitem[\protect\astroncite{{Halzen} et~al.}{1995}]{Halzen95}
{Halzen}, F., {Vazquez}, R.~A., {Stanev}, T., and {Vankov}, H.~P., 1995,
\newblock {Astroparticle Physics} {3}, 151

\bibitem[\protect\astroncite{{Hayashida}}{1996}]{AGASA_pairs}
{Hayashida}, N. e.~a., 1996,
\newblock {Phys. Rev. Lett.} {77}, 1000

\bibitem[\protect\astroncite{{Hillas}}{1998}]{Hillas98}
{Hillas}, M., 1998,
\newblock {Nature} {395}, 15

\bibitem[\protect\astroncite{{Huchra} et~al.}{1992}]{Huchra_ZCAT}
{Huchra}, J.~P., {Geller}, M.~J., {Clemens}, C.~M., {Tokarz}, S.~P., and
  {Michel}, A., 1992,
\newblock {Bulletin d'Information du Centre de Donnees Stellaires} {41}, 31

\bibitem[\protect\astroncite{{Ishwara-Chandra} and
  {Saikia}}{1999}]{1999MNRAS.309..100I}
{Ishwara-Chandra}, C.~H. and {Saikia}, D.~J., 1999,
\newblock {\mnras} {309}, 100

\bibitem[\protect\astroncite{{Komissarov} and
  {Gubanov}}{1994}]{1994A&A...285...27K}
{Komissarov}, S.~S. and {Gubanov}, A.~G., 1994,
\newblock {\aap} {285}, 27

\bibitem[\protect\astroncite{{Kronberg}}{1994a}]{Kromberg94}
{Kronberg}, P.~P., 1994a,
\newblock {Rep. Prog. Phys.} {57}, 325

\bibitem[\protect\astroncite{{Kronberg}}{1994b}]{1994RPPh...57..325K}
{Kronberg}, P.~P., 1994b,
\newblock {Reports of Progress in Physics} {57}, 325

\bibitem[\protect\astroncite{{Kronberg} et~al.}{1999}]{1999ApJ...511...56K}
{Kronberg}, P.~P., {Lesch}, H., and {Hopp}, U., 1999,
\newblock {\apj} {511}, 56

\bibitem[\protect\astroncite{{Linsley}}{1975}]{Linsley_1st_harmonic}
{Linsley}, J., 1975,
\newblock {Phys. Rev. Lett.} {34}, 1530

\bibitem[\protect\astroncite{{Linsley}}{1980}]{VR_4e19}
{Linsley}, J., 1980,
\newblock {Catalogue of Highest Energy Cosmic Rays(ed. M Wada) WDC C2 for CR,
  IPCR, Tokyo} {1}, 61

\bibitem[\protect\astroncite{{Malkan} and
  {Stecker}}{1998}]{Malkan_and_Stecker_98}
{Malkan}, M.~A. and {Stecker}, F.~W., 1998,
\newblock {Ap. J.} {496}, 13

\bibitem[\protect\astroncite{{Medina Tanco}}{1997}]{GMT_Durban_IGMF97}
{Medina Tanco}, G.~A., 1997,
\newblock in {25th ICRC}, Vol.~4, p. 477

\bibitem[\protect\astroncite{{Medina Tanco}}{1998a}]{GMT_APJL98_cluster}
{Medina Tanco}, G.~A., 1998a,
\newblock {\apjl} {495}, L71

\bibitem[\protect\astroncite{{Medina Tanco}}{1998b}]{GMT_APJL98_IGMF}
{Medina Tanco}, G.~A., 1998b,
\newblock {\apjl} {505}, L79

\bibitem[\protect\astroncite{{Medina Tanco}}{2000a}]{UHECRiso00}
{Medina Tanco}, G.~A., 2000a,
\newblock {Ap. J. (accepted)}

\bibitem[\protect\astroncite{{Medina Tanco}}{2000b}]{GMT_isotropy00}
{Medina Tanco}, G.~A., 2000b,
\newblock {submitted}

\bibitem[\protect\astroncite{{Medina Tanco} and
  {Watson}}{1999}]{GMT_AAW_DM_halo}
{Medina Tanco}, G.~A. and {Watson}, A.~A., 1999,
\newblock {Astroparticle Physics} {12}, 25

\bibitem[\protect\astroncite{{Peacock}}{1999}]{1999coph.book.....P}
{Peacock}, J.~A., 1999,
\newblock {Cosmological physics},
\newblock Cosmological physics. Publisher: Cambridge, UK: Cambridge University
  Press, 1999. ISBN: 0521422701

\bibitem[\protect\astroncite{{Puget} et~al.}{1976}]{Puget76}
{Puget}, J.~L., {Stecker}, F.~W., and {Bredekamp}, J.~H., 1976,
\newblock {Ap. J.} {205}, 638

\bibitem[\protect\astroncite{{Rawlings} and
  {Saunders}}{1991}]{1991Natur.349..138R}
{Rawlings}, S. and {Saunders}, R., 1991,
\newblock {\nat} {349}, 138

\bibitem[\protect\astroncite{{Reid} and {Watson}}{1980}]{HP_4e19}
{Reid}, R. J.~O. and {Watson}, A.~A., 1980,
\newblock {Catalogue of Highest Energy Cosmic Rays (ed. M Wada) WDC C2 for CR,
  IPCR, Tokyo} {1}, 61

\bibitem[\protect\astroncite{{Roettiger} et~al.}{1999}]{1999ApJ...518..594R}
{Roettiger}, K., {Stone}, J.~M., and {Burns}, J.~O., 1999,
\newblock {\apj} {518}, 594

\bibitem[\protect\astroncite{{Ryu} et~al.}{1998a}]{1998A&A...335...19R}
{Ryu}, D., {Kang}, H., and {Biermann}, P.~L., 1998a,
\newblock {\aap} {335}, 19

\bibitem[\protect\astroncite{{Ryu} et~al.}{1998b}]{Ryu98}
{Ryu}, D., {Kang}, H., and {Biermann}, P.~L., 1998b,
\newblock {\aap} {335}, 19

\bibitem[\protect\astroncite{{Slee} and {Roy}}{1998}]{1998MNRAS.297L..86S}
{Slee}, O.~B. and {Roy}, A.~L., 1998,
\newblock {\mnras} {297}, L86

\bibitem[\protect\astroncite{{Soltan}}{1982}]{1982MNRAS.200..115S}
{Soltan}, A., 1982,
\newblock {\mnras} {200}, 115

\bibitem[\protect\astroncite{{Stecker}}{1998}]{Stecker98}
{Stecker}, F.~W., 1998,
\newblock {Workshop on observing GAS from Space, J. F. Krizmanic et al. (eds.)}
  pp 212--215

\bibitem[\protect\astroncite{{Takeda}}{1999}]{AG_1e19}
{Takeda}, M. e.~a., 1999,
\newblock {Astrophys. J.} {522}, 225

\bibitem[\protect\astroncite{{Vallee}}{1997}]{Vallee97}
{Vallee}, J.~P., 1997,
\newblock {Fundamentals of Cosmic Physics} {19}, 1

\bibitem[\protect\astroncite{{Venturi} et~al.}{1998}]{1998MNRAS.298.1113V}
{Venturi}, T., {Bardelli}, S., {Morganti}, R., and {Hunstead}, R.~W., 1998,
\newblock {\mnras} {298}, 1113

\bibitem[\protect\astroncite{{Waxman} et~al.}{1997}]{Waxman97}
{Waxman}, E., {Fisher}, K.~B., and {Piran}, T., 1997,
\newblock {\apj} {483}, 1

\bibitem[\protect\astroncite{{Zatsepin} and {Kuzmin}}{1966}]{GZK_ZK66}
{Zatsepin}, G.~T. and {Kuzmin}, V.~A., 1966,
\newblock {Sov. Phys. JETP Lett.} {4}, 78

\end{thebibliography}
\bibliographystyle{aabib99}

\newpage

\noindent
{\bf \large Figure Captions:}\\[2em]

\noindent {Fig. \ref{fig1}. Energy distribution of radio ghosts
multiplied by $E_\gh^2$ for better display. The lines are labeled
with $b_\nu$, the exponent of the radio luminosity-jetpower
correlation. The solid lines are derived from the {\it pure
luminosity evolution} (PLE) radio luminosity function of Dunlop \&
Peacock (1990). The dashed lines from their {\it radio luminosity
function No. 2} from the {\it MEAN-z data} (RLF2). This latter
luminosity function has an exponential cutoff at luminosities of
$10^{28}$ Watt Hz$^{-1}$. The curves are plotted for the range of
radio luminosities between $10^{23}$ to $10^{29}$ Watt Hz$^{-1}$.
A horizontal line in this diagram corresponds to $\alpha_\gh = 2$.
E.g. the PLE with $b_\nu = 1.0$ has a low energy index of
$\alpha_\gh = 1.75$.}\\[2em]

\noindent
{Fig. \ref{fig2}.  Example of a possible internal radio
ghost magnetic field topology.  The vectorial representation of
the component of the magnetic field parallel to the plane of a
slice, cut through the middle of the ghost, is shown. The length
of the arrows scales linearly with magnetic field intensity. The
magnetic field is tangent to the surface of the ghost. The
amplitude of the fluctuations obeys a power law power spectrum of
spectral index $\xi = 4$. See text for further details. }\\[2em]

\noindent {Fig. \ref{fig3}.  (a) Angular distribution of scattered
UHECR for ghosts of different radius (in units of proton
gyroradius) and values of $\xi$. The incident particles have a
power law spectrum, $dN/dE \propto E^{-3}$ for E $> 4 \times
10^{19}$ eV. For comparison, the thick line shows an isotropic
angular distribution. Forward scattering becomes dominant only as
the ghost radius becomes comparable to the UHECR gyroradius. (b)
Same as (a) but for protons at a fixed energy, E$=10^{20}$ eV. The
same ghosts are used as in (a) and so their radii are down by a
factor of approximately two in units of particle
gyroradius.}\\[2em]

\noindent {Fig. \ref{fig4}.  UHECR propagation without radio
ghosts.  Aitoff projection of the arrival probability density.
Galactic coordinates are used with the antigalactic center at the
center of the figure. UHECR sources are distributed according to
nearby luminous matter (CfA catalog). Data points correspond to
cosmic rays observed by AGASA, Haverah Park, Yakutsk and Volcano
Ranch with $E > 4 \times 10^{19}$ eV. Clearly, the observations
are more isotropic than what should be expected from the model.
}\\[2em]

\noindent
{Fig. \ref{fig5}.
UHECR received at Earth for $\alpha_{gh} = 2.0$, $b_{gh} = 2.0$
and sources distributed as luminous matter.
}\\[2em]

\noindent
{Fig. \ref{fig6}
UHECR received at Earth for $\alpha_{gh} = 2.0$, $b_{gh} = 1.0$
and sources distributed as luminous matter. }\\[2em]

\noindent
{Fig. \ref{fig7}
UHECR received at Earth for $\alpha_{gh} = 2.0$, $b_{gh} = 0.5$
and sources distributed as luminous matter.
(a) UHECR that
encounter no ghost from source to Earth; (b) UHECR
that underwent at least one scattering off a RG.
Comparing this figure with Fig. \ref{fig4}, \ref{fig5} and \ref{fig6}, it
can be seen that ghosts must have a more extended
distribution than the sources ($b_{gh} < 0.5$) in
order to isotropize the UHECR flux at Earth.
A tolerable (3-fold) increase in average ghosts energy density
goes in the right direction but is not a determinant factor
in itself.} \\[2em]

\noindent
{Fig. \ref{fig8}
Amplitude of the first harmonic for UHECR with
$E > 4 \times 10^{19}$ eV for $\alpha = 2.0$ and $b_{gh} = 0.5$.
1000 independent samples of 50 UHECR each were selected
from the distribution in Fig. \ref{fig7}c with the same sensitivity
in declination as AGASA's. The amplitudes of the first
harmonic obtained for these samples are shown in the histogram.
As a comparison, the amplitudes obtained by AGASA, Haverah Park,
Yakutsk, and Volcano Ranch \cite{GMT_AAW_DM_halo}
are also shown. It can be seen that, in order to isotropize the
UHECR flux, the condition $b_{gh} {<} 0.5$ must be fulfilled.
That is, ghosts should be much more smoothly distributed in space
than the galaxies that originated them.
}\\[2em]

\noindent {Fig. \ref{fig9} UHECR mean free path [Mpc] as a
function of $b_\gh$, the bias between galaxies and ghost
distributions, and gas density relative to the reference level.
Ghosts are able to isotropize the UHECR flux only for $b_{gh}
\stackrel{~}{<} 0.5$, for which the mean free path is smaller than
a few Mpc both at voids and structures.}\\[2em]

\noindent
{Fig. \ref{fig10}
UHECR received at Earth for $\alpha_{gh} = 2.0$, $b_{gh} = 2.0$
and sources distributed as $\Lambda$CDM. Supergalactic coordinates are
used in this figure. The strong excess in the arrival direction
distribution corresponds to the Virgo cluster.}\\[2em]

\noindent
{Fig. \ref{fig11}
UHECR received at Earth for $\alpha_{gh} = 2.0$, $b_{gh} = 0.5$
and sources distributed as $\Lambda$CDM. Again, isotropization
is obtained only for a ghost population that is much more
extended than the UHECR sources. Supergalactic coordinates are
used in this figure as in Fig. \ref{fig10}.}\\[2em]

\noindent {Fig. \ref{fig12} Distribution of injected magnetic
field strength as a function of background density, in the case
that all radio ghosts completely mix with the ambient medium. The
volume fractions of the different density regimes can be seen in
Fig. \ref{fig13}.}\\[2em]

\noindent {Fig. \ref{fig13} The volume fractions of the different
density regimes ($dV/d\log (n_{\rm gas} / n_{\rm gas,0})/\int dV$)
computed from the simulated dark matter distribution. The median
density is marked. A count in cell box with size of $1.5$ Mpc was
used.}\\[2em]

\figyn{
\newpage
\begin{figure*}[thb]
\begin{center}
\figOyn{\psfig{figure=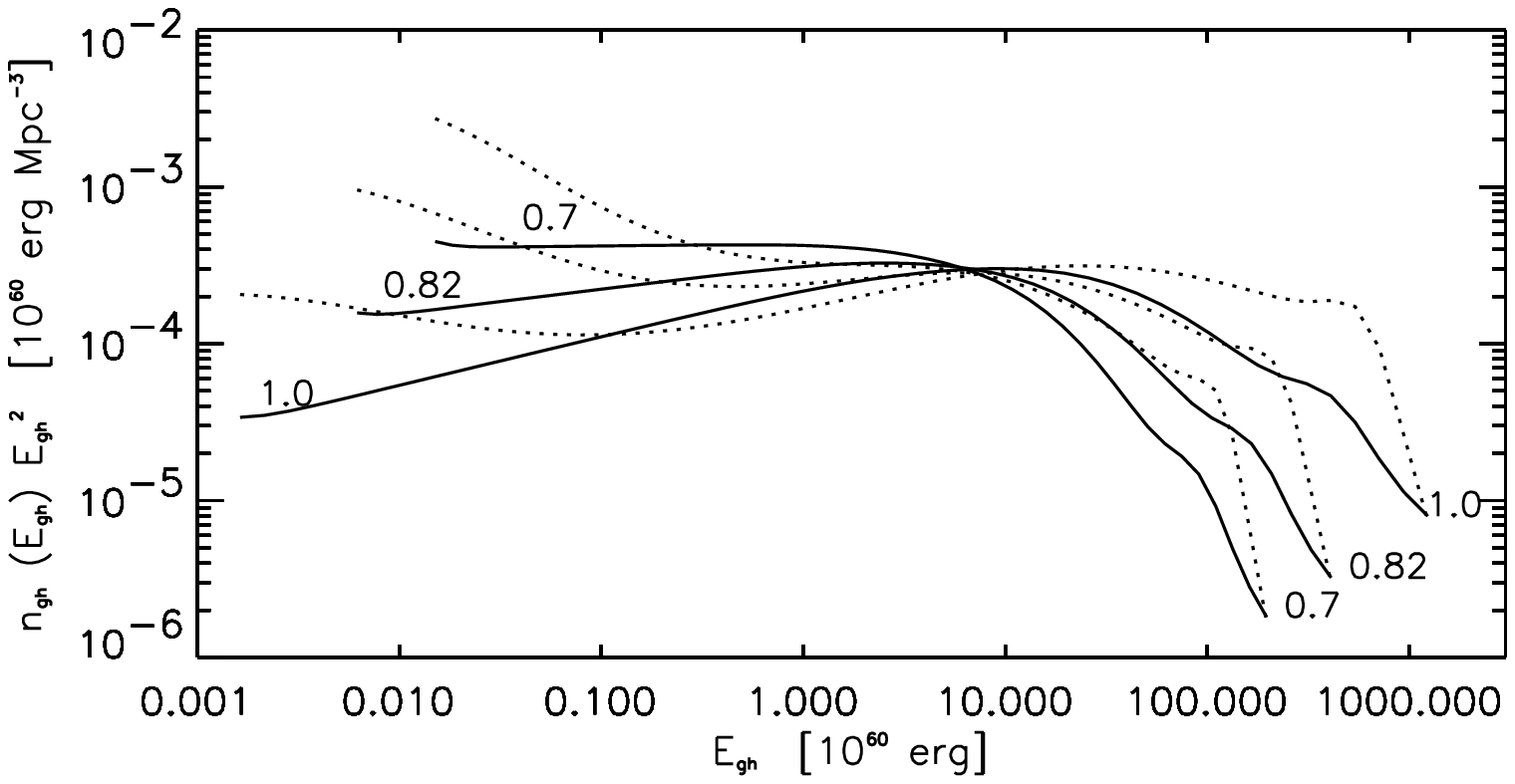,width=0.9\textwidth,angle=0}}
\figPSyn{\psfig{figure=figure1.ps,width=0.9\textwidth,angle=0}}
\figpsyn{\psfig{figure=figu1s.ps,width=0.9\textwidth,angle=0}}
\caption[]{\label{fig1}}
\end{center}
\end{figure*}
}

\figyn{\figOyn{\newpage}
\begin{figure}[!hbt]
\figOyn{\centerline{\psfig{figure=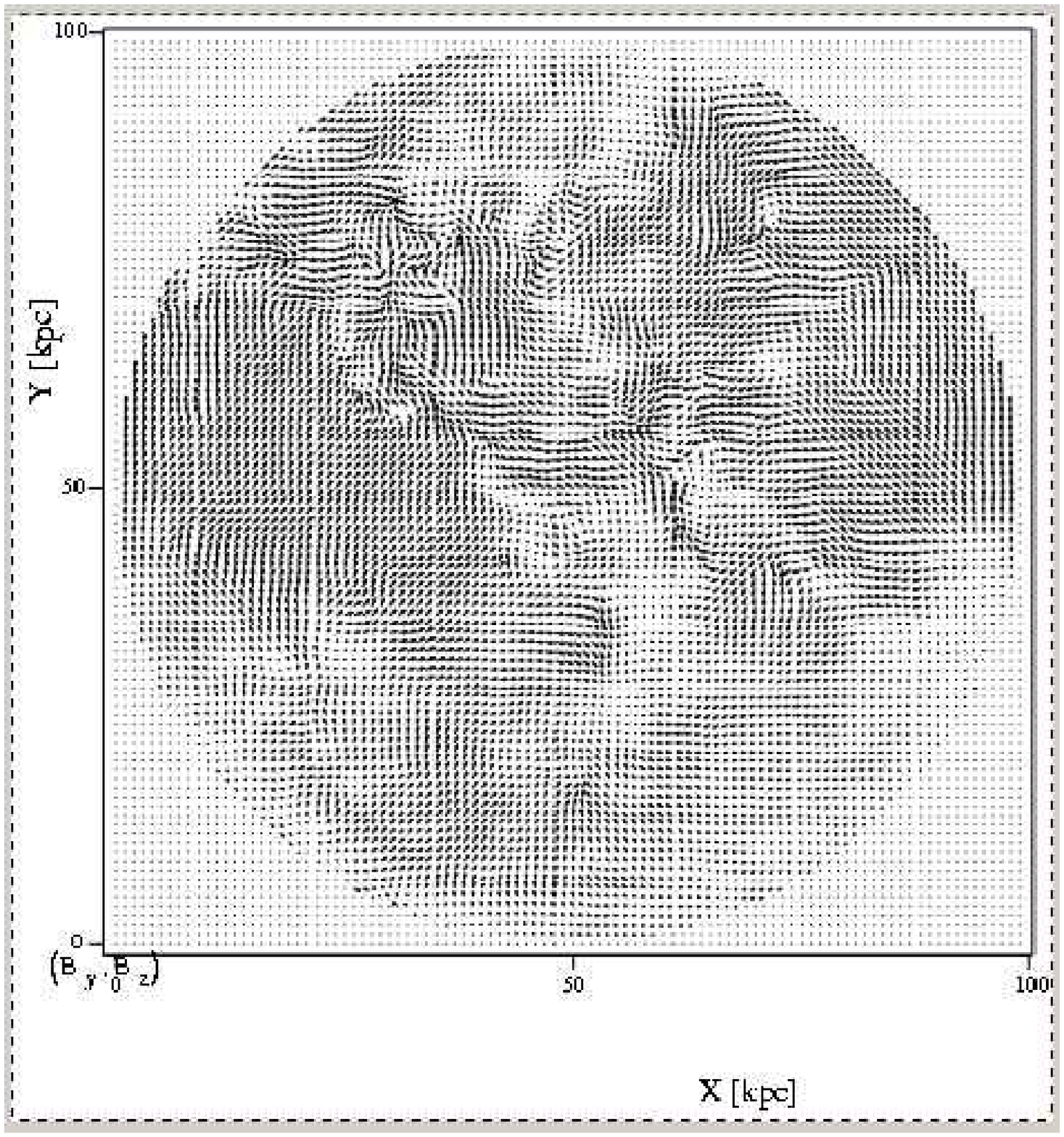,width=14cm}}}
\figPSyn{\centerline{\psfig{figure=figu2.ps,width=12cm}}}
\figpsyn{\centerline{\psfig{figure=figu2s.ps,width=12cm}}}
\caption[]{\label{fig2}}
\end{figure}
}

\figyn{\figOyn{\newpage}
\begin{figure}[!hbt]
\figOyn{\centerline{\psfig{figure=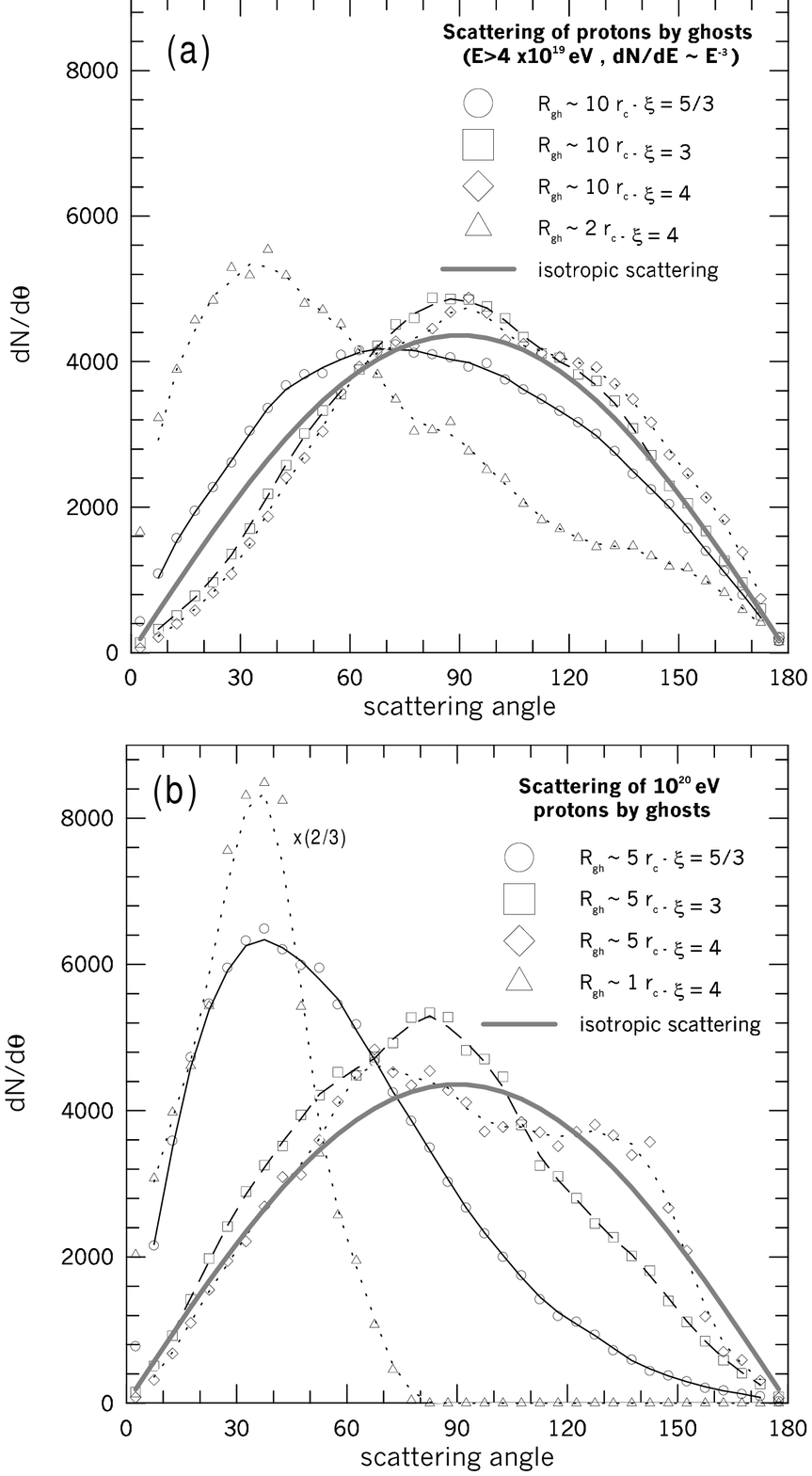,width=16.5cm}}}
\figPSyn{\centerline{\psfig{figure=figu3.ps,width=12cm}}}
\figpsyn{\centerline{\psfig{figure=figu3s.ps,width=12cm}}}
\caption[]{\label{fig3}}
\end{figure}
}

\figOyn{\newpage}

\figyn{
\begin{figure}[!hbt]
\figOyn{\centerline{\psfig{figure=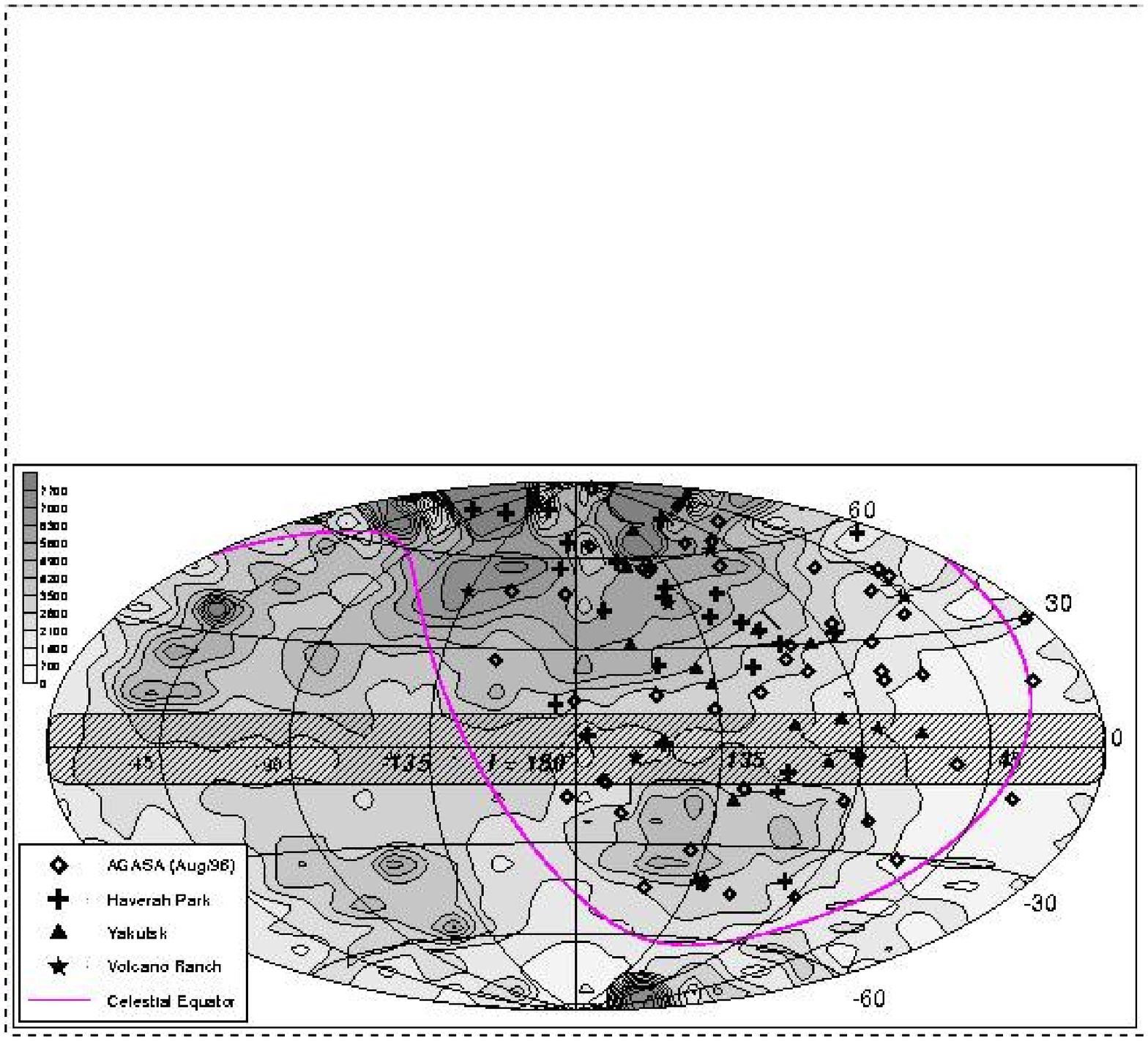,width=15cm}}}
\figPSyn{\centerline{\psfig{figure=figu4.ps,width=17cm}}}
\figpsyn{\centerline{\psfig{figure=figu4s.ps,width=17cm}}}
\caption[]{\label{fig4}}
\end{figure}
}

\figOyn{\newpage}

\figOyn{\newpage} \figyn{
\begin{figure}[!tb]
\figOyn{\centerline{\psfig{figure=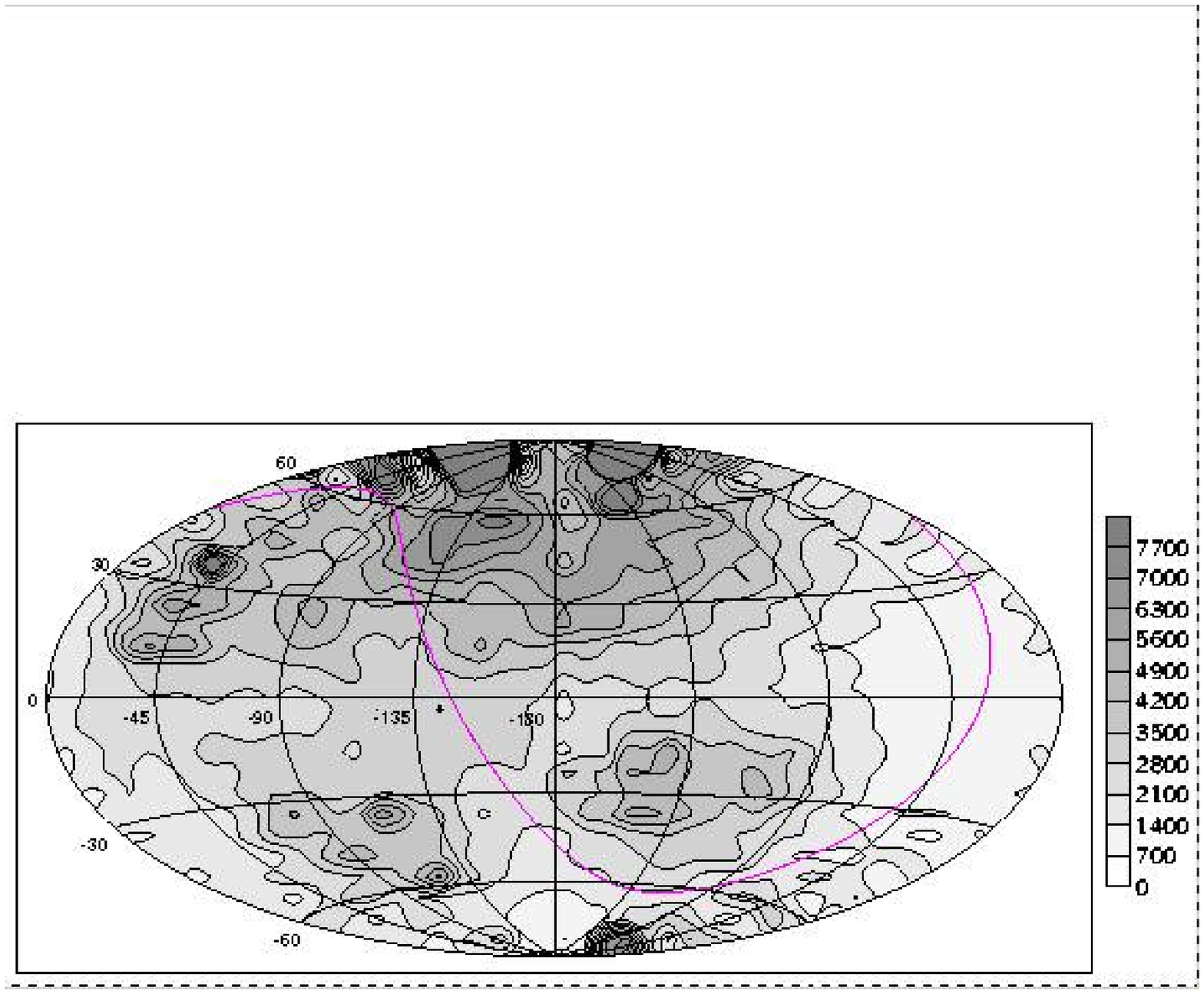,width=15cm}}}
\figPSyn{\centerline{\psfig{figure=figu5.ps,width=15cm}}}
\figpsyn{\centerline{\psfig{figure=figu5s.ps,width=15cm}}}
\caption[]{\label{fig5}}
\end{figure}
}

\figOyn{\newpage}

\figyn{
\begin{figure}[t]
\figOyn{\centerline{\psfig{figure=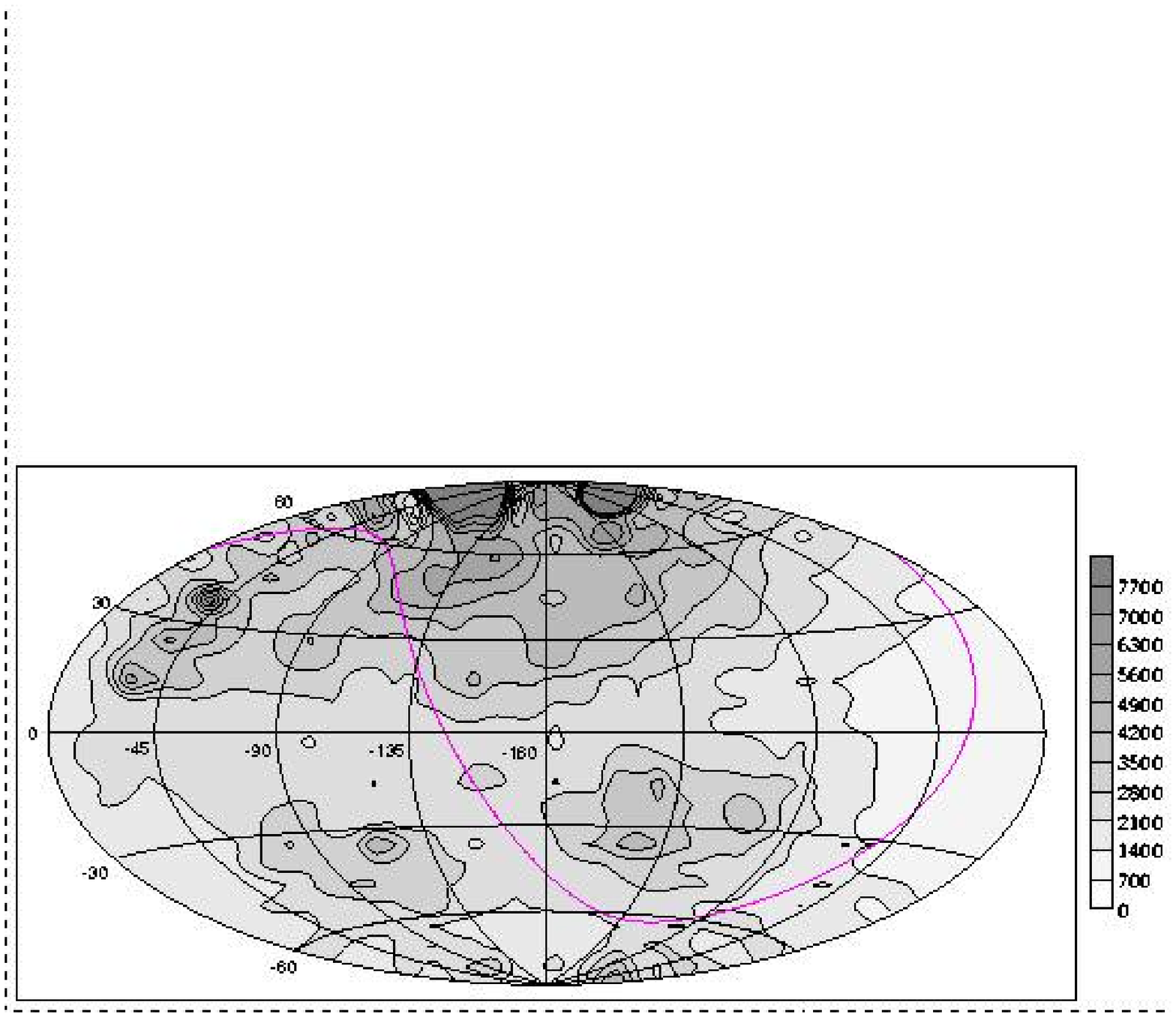,width=15cm}}}
\figPSyn{\centerline{\psfig{figure=figu6.ps,width=15cm}}}
\figpsyn{\centerline{\psfig{figure=figu6s.ps,width=15cm}}}
\caption[]{\label{fig6}}
\end{figure}
}

\figOyn{\newpage}

\figyn{
\begin{figure}[t]
\figOyn{\centerline{\psfig{figure=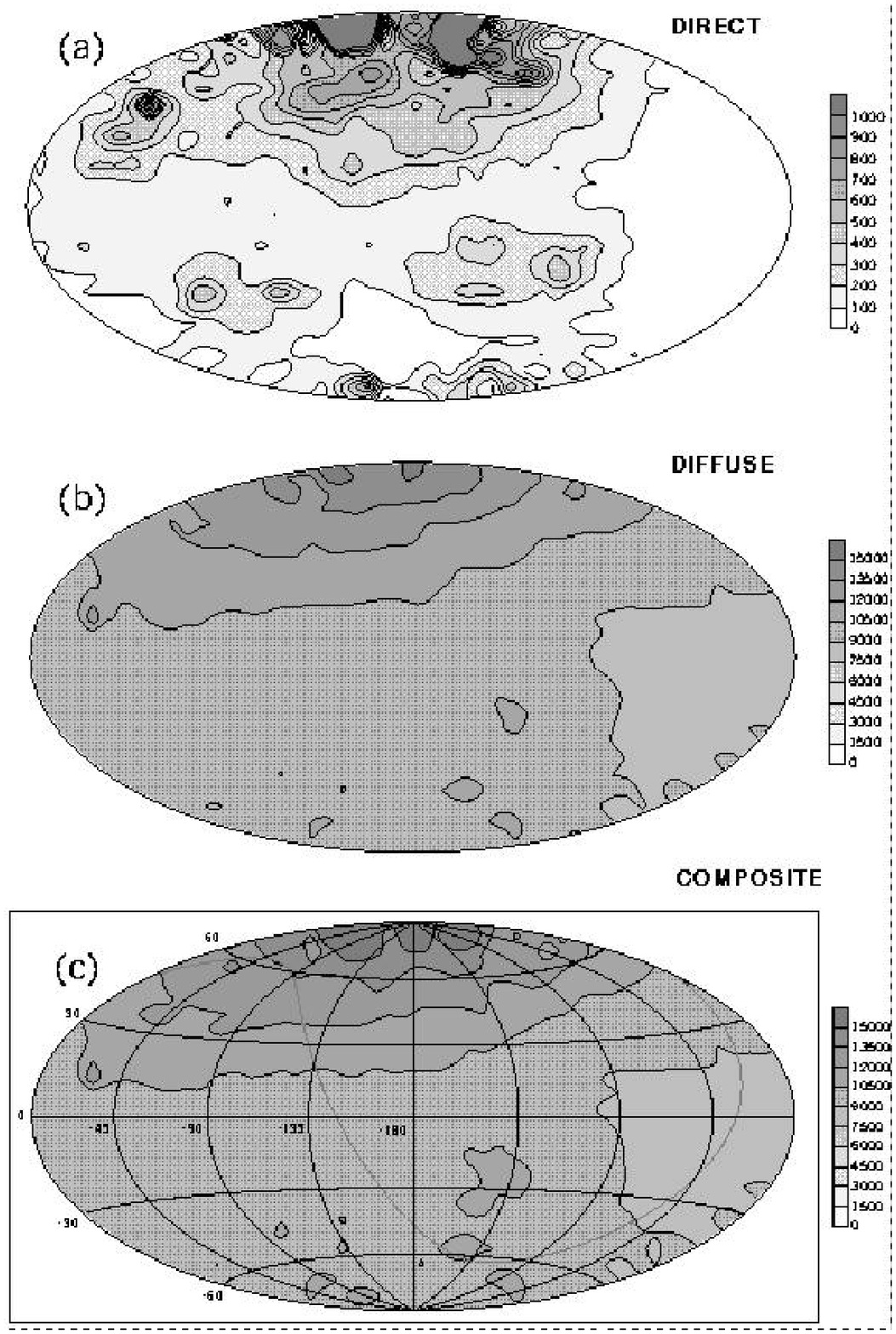,width=14cm}}}
\figPSyn{\centerline{\psfig{figure=figu7abc.ps,width=14cm}}}
\figpsyn{\centerline{\psfig{figure=figu7s.ps,width=14cm}}}
\caption[]{\label{fig7}}
\end{figure}
}

\figOyn{\newpage}

\figyn{
\begin{figure}[thb]
\figOyn{\centerline{\psfig{figure=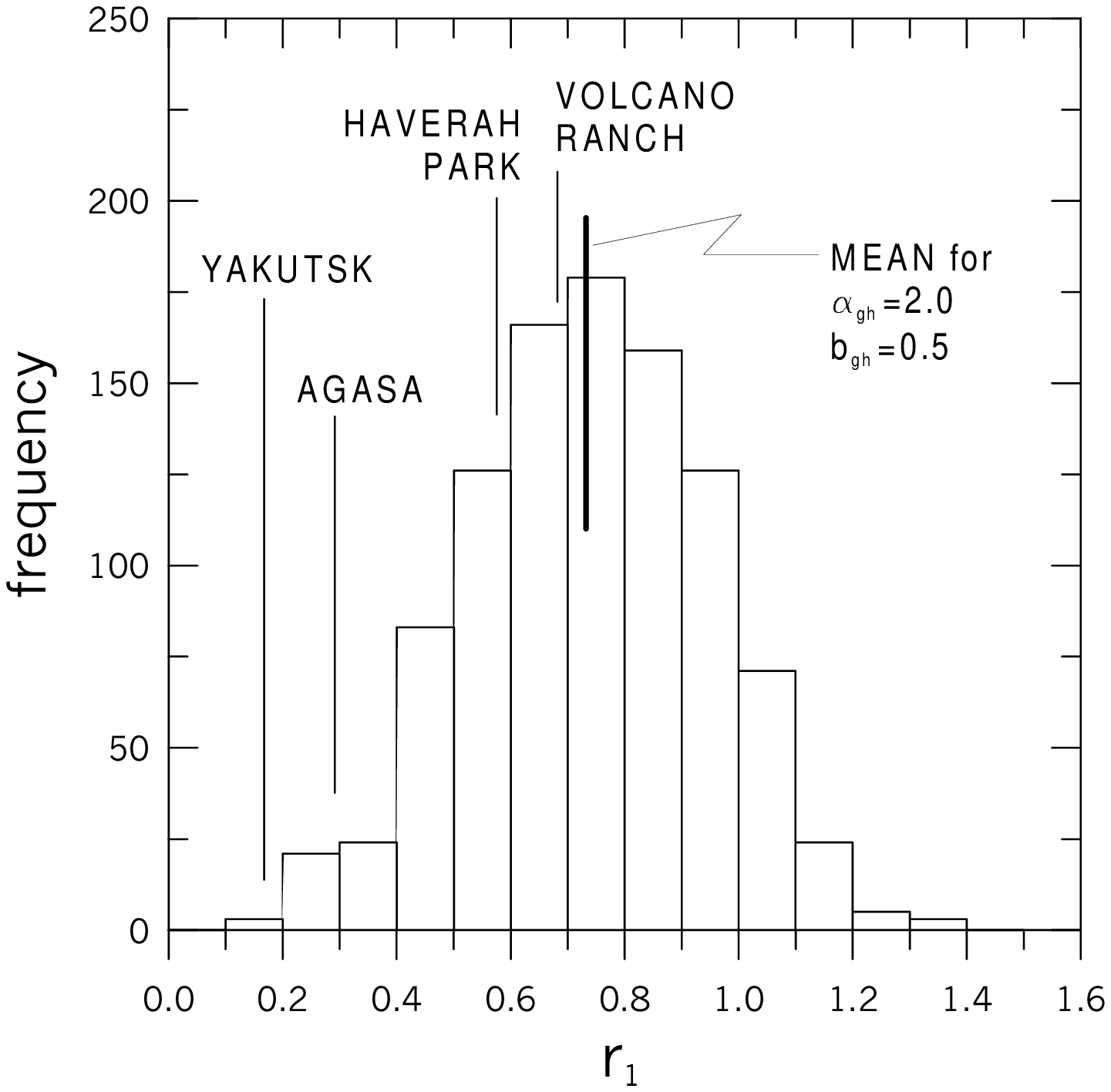,width=12cm}}}
\figPSyn{\centerline{\psfig{figure=figu8.ps,width=12cm}}}
\figpsyn{\centerline{\psfig{figure=figu8s.ps,width=12cm}}}
\caption[]{\label{fig8}}
\end{figure}
}

\figOyn{\newpage}

\figyn{
\begin{figure}[thb]
\figOyn{\centerline{\psfig{figure=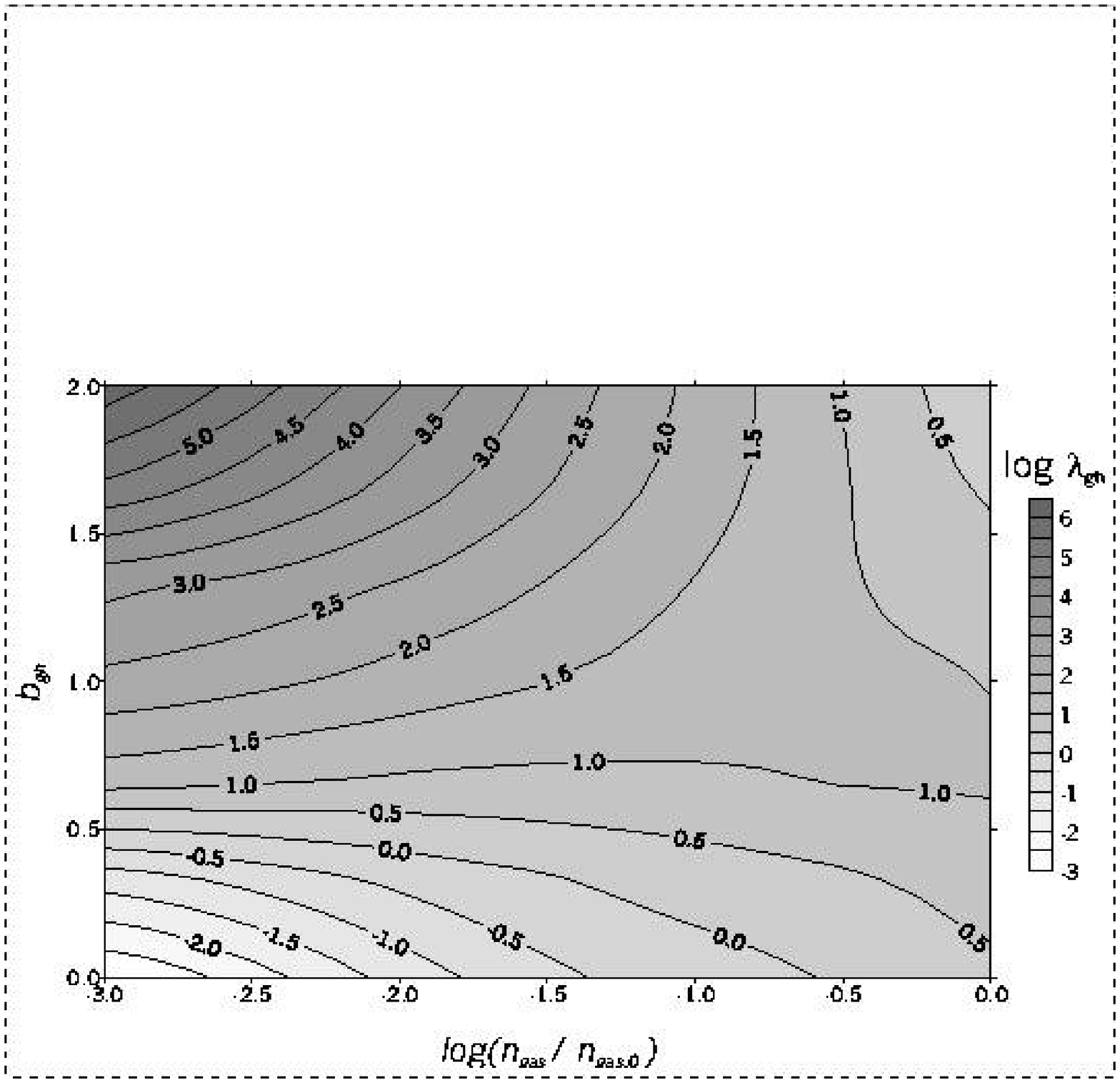,width=12cm}}}
\figPSyn{\centerline{\psfig{figure=figu9.ps,width=12cm}}}
\figpsyn{\centerline{\psfig{figure=figu9s.ps,width=12cm}}}
\caption[]{\label{fig9}}
\end{figure}
}

\figOyn{\newpage}

\figyn{
\begin{figure}[t]
\figOyn{\centerline{\psfig{figure=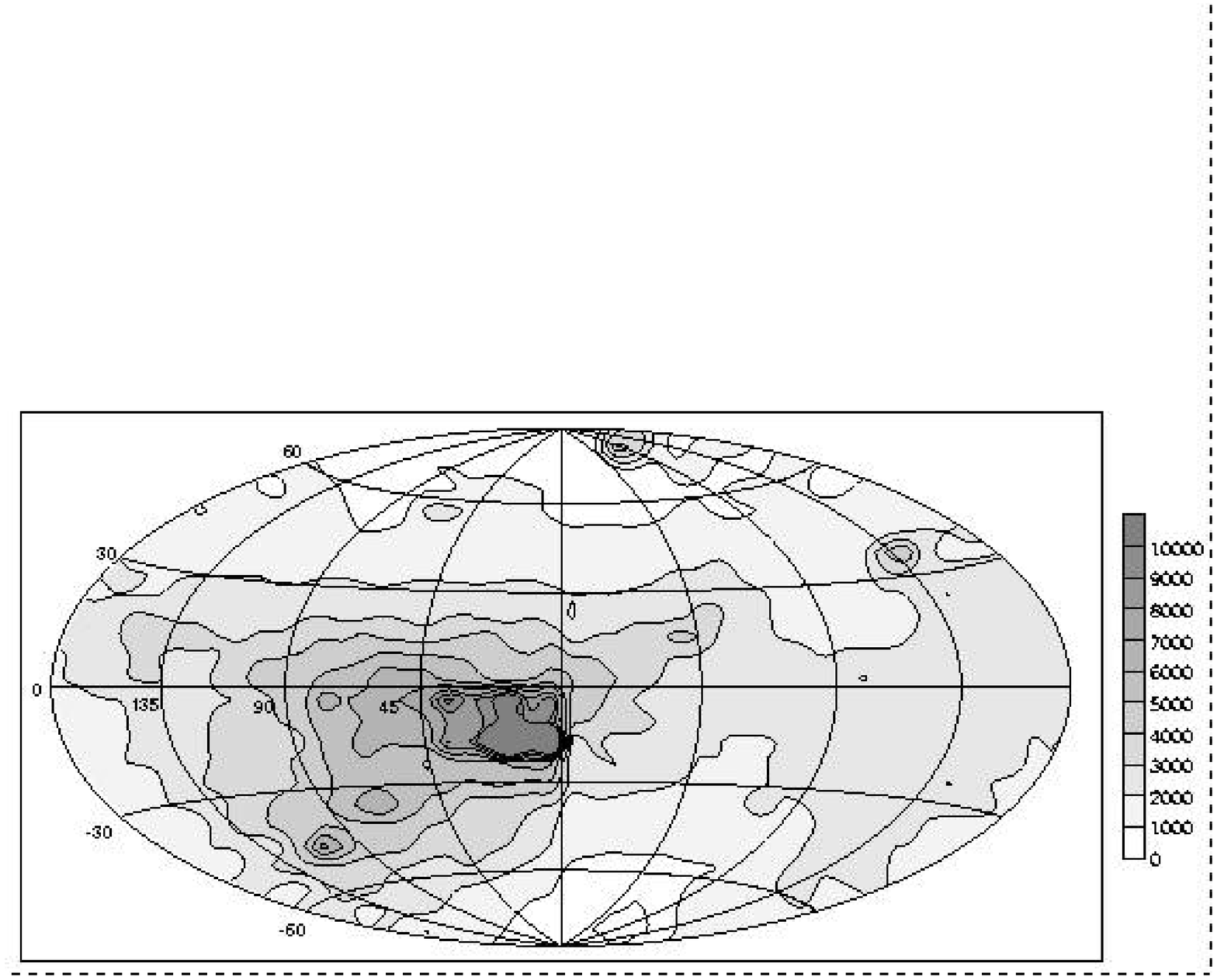,width=15cm}}}
\figPSyn{\centerline{\psfig{figure=figu10.ps,width=15cm}}}
\figpsyn{\centerline{\psfig{figure=figu10s.ps,width=15cm}}}
\caption[]{\label{fig10}}
\end{figure}
}

\figOyn{\newpage}

\figyn{
\begin{figure}[t]
\figOyn{\centerline{\psfig{figure=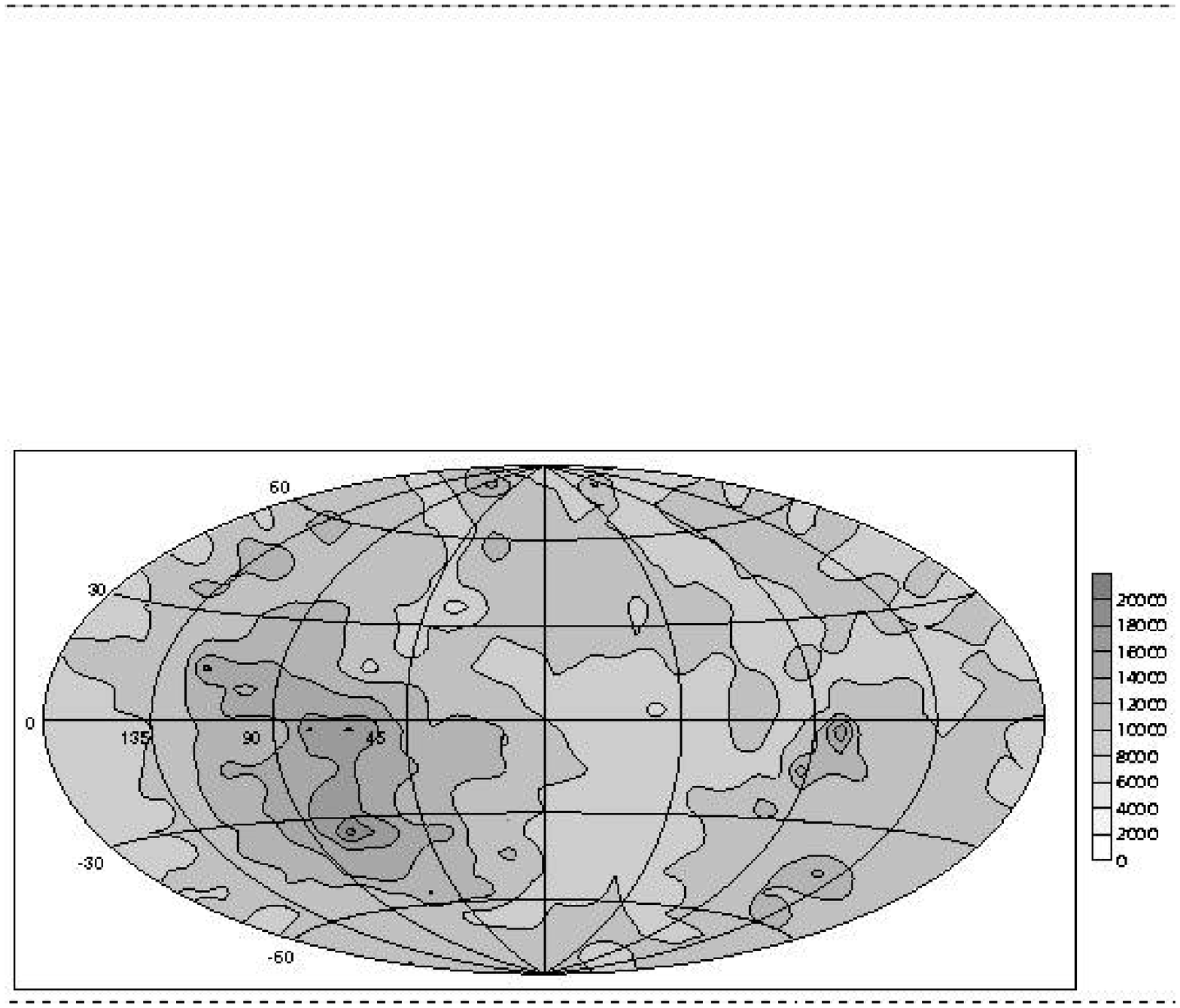,width=15cm}}}
\figPSyn{\centerline{\psfig{figure=figu11.ps,width=15cm}}}
\figpsyn{\centerline{\psfig{figure=figu11s.ps,width=15cm}}}
\caption[]{\label{fig11}}
\end{figure}
}

\figOyn{\newpage}

\figyn{
\begin{figure}[t]
\figOyn{\centerline{\psfig{figure=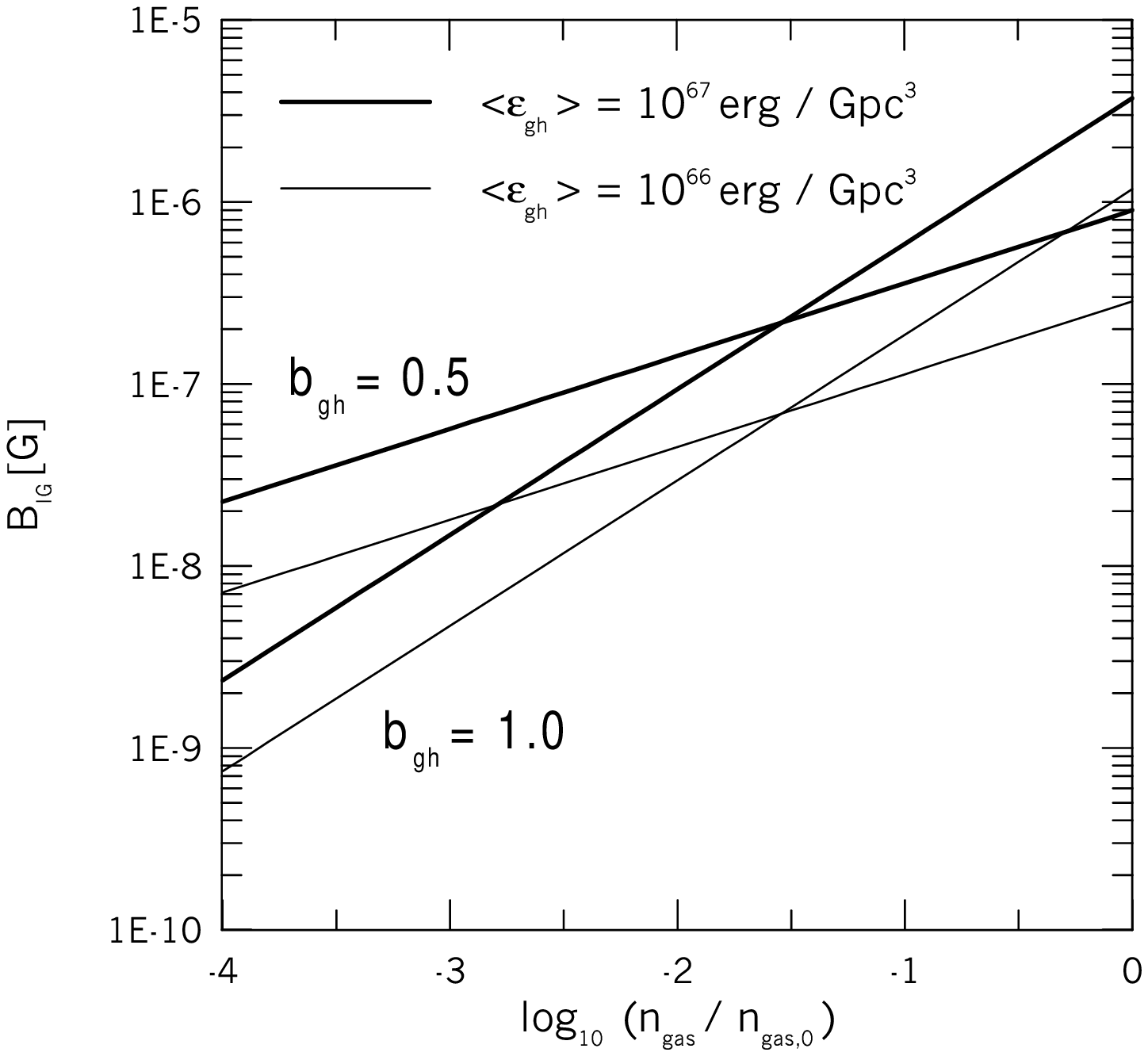,width=15cm}}}
\figPSyn{\centerline{\psfig{figure=figure12.ps,width=15cm}}}
\figpsyn{\centerline{\psfig{figure=figure12.ps,width=15cm}}}
\caption[]{\label{fig12}}
\end{figure}
}

\figOyn{\newpage}

\figyn{
\begin{figure}[t]
\figOyn{\centerline{\psfig{figure=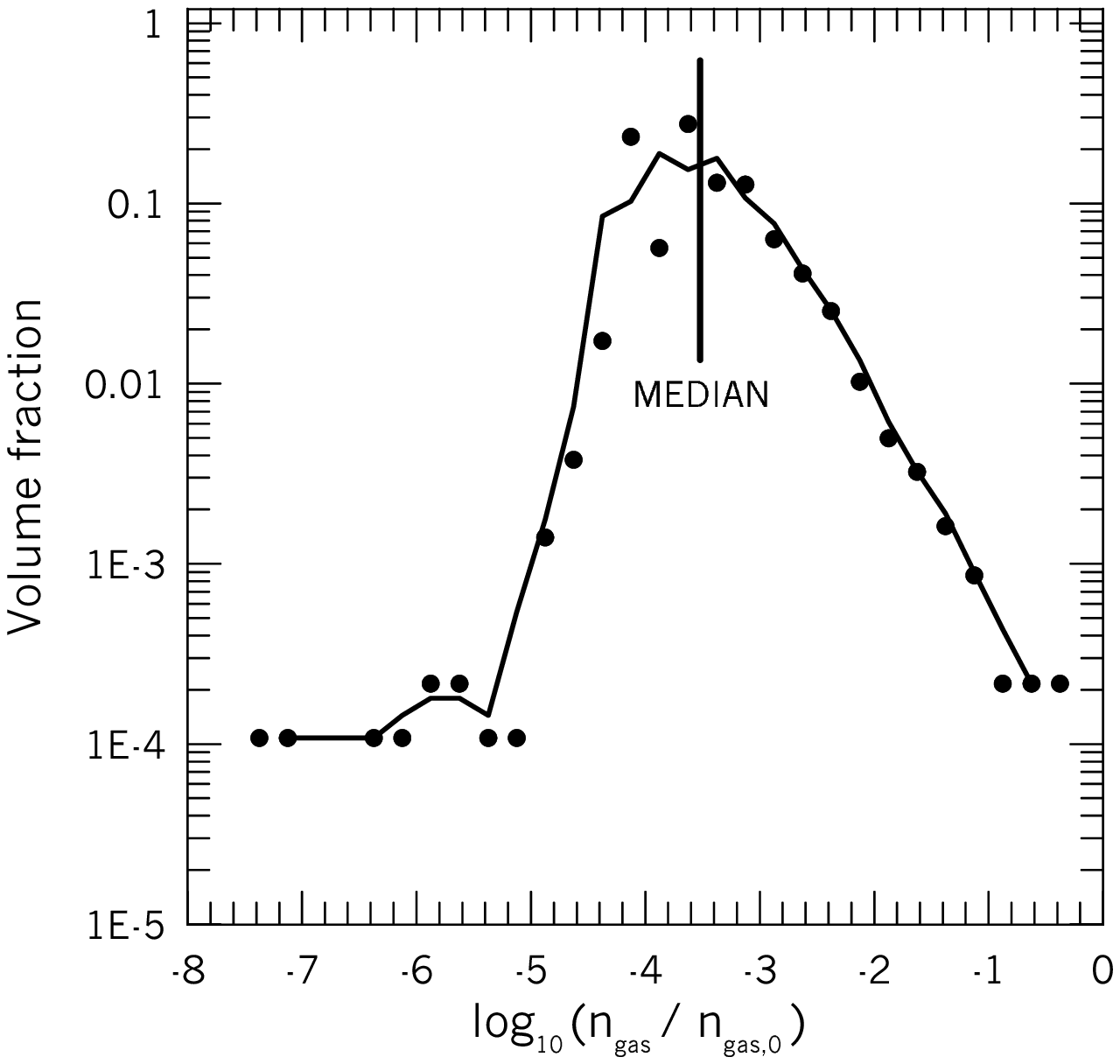,width=15cm}}}
\figPSyn{\centerline{\psfig{figure=figure13.ps,width=15cm}}}
\figpsyn{\centerline{\psfig{figure=figure13.ps,width=15cm}}}
\caption[]{\label{fig13}}
\end{figure}
}
\end{document}